\documentclass[10pt]{iopart}
\expandafter\let\csname equation*\endcsname\relax
\expandafter\let\csname endequation*\endcsname\relax
\usepackage{amsmath}
\usepackage{amsbsy,amsfonts,amssymb,graphics,epsfig,float,mathrsfs,amsthm,braket}
\usepackage{rotating,booktabs,array,tabularx}
\usepackage{ifsym}
\usepackage{psfrag}
\usepackage[normalem]{ulem}
\usepackage{color}
\usepackage[11pt]{moresize}
\usepackage{longtable,boldline}
\usepackage{appendix}
\usepackage{enumitem}

\newcommand \beq{\begin{equation}}
\newcommand \eeq{\end{equation}}
\newcommand \beqn{\begin{equation*}}
\newcommand \eeqn{\end{equation*}}

\newcommand{\Wf}{W_{\mathrm{fold}}}
\newcommand{\Uf}{U_{\mathrm{fold}}}
\newcommand{\tauf}{\tau_{\mathrm{fold}}}
\newcommand{\lamf}{\lambda_{\mathrm{fold}}}
\newcommand{\cS}{\mathcal{S}}
\newcommand{\Wt}{\tilde{W}}
\newcommand{\Ut}{\tilde{U}}
\newcommand{\taut}{\tilde{\tau}}

\newcommand{\id}{\mathrm{d}} 
\renewcommand{\d}[2]{\frac{\id #1}{\id #2}} 
\newcommand{\dd}[2]{\frac{\id^2 #1}{\id #2^2}} 
\newcommand{\df}[2]{\frac{\id^4 #1}{\id #2^4}} 

\newcommand{\pd}[2]{\frac{\partial #1}{\partial #2}} 
\newcommand{\pdd}[2]{\frac{\partial^2 #1}{\partial #2^2}} 
\newcommand{\pdf}[2]{\frac{\partial^4 #1}{\partial #2^4}} 

   \newcommand{\citep}[1]{\cite{#1}}
\begin{document}

\title{Pull-in dynamics of overdamped microbeams}
\footnote{This is an author-created, un-copyedited version of an article accepted for publication/published in the Journal of Micromechanics and Microengineering. IOP Publishing Ltd is not responsible for any errors or omissions in this version of the manuscript or any version derived from it. The Version of Record is available online at  https://doi.org/10.1088/1361-6439/aad72f.}
\author{Michael Gomez, Dominic Vella, Derek E.~Moulton}
\address{Mathematical Institute, University of Oxford, Woodstock Rd, Oxford, OX2 6GG, UK}
\ead{moulton@maths.ox.ac.uk}

\begin{abstract}
We study the dynamics of MEMS microbeams undergoing electrostatic pull-in. At DC voltages close to the pull-in voltage, experiments and numerical simulations have reported `bottleneck' behaviour in which the transient dynamics slow down considerably. This slowing down is highly sensitive to external forces, and so has widespread potential for applications that use pull-in time as a sensing mechanism, including high-resolution accelerometers and pressure sensors. Previously, the bottleneck phenomenon has only been understood using lumped mass-spring models that do not account for effects such as variable residual stress and different boundary conditions. We extend these studies to incorporate the beam geometry, developing an asymptotic method to analyse the pull-in dynamics. We attribute bottleneck behaviour to critical slowing down near the pull-in transition, and we obtain a simple expression for the pull-in time in terms of the beam parameters and external damping coefficient. This expression is found to agree well with previous experiments and numerical simulations that incorporate more realistic models of squeeze film damping, and so provides a useful design rule for sensing applications. We also consider the accuracy of a single-mode approximation of the microbeam equations --- an approach that is commonly used to make analytical progress, without systematic investigation of its accuracy. By comparing to our bottleneck analysis, we identify the factors that control the error of this approach, and we demonstrate that this error can indeed be very small.
\end{abstract}

\vspace{2pc}
\noindent{\it Keywords}: \\Electrostatic pull-in,\\Microbeam,\\Switching time,\\MEMS\\
\linebreak
\maketitle
\ioptwocol

\section{Introduction}
\label{sec:intro}

\begin{figure}
\centering
\includegraphics[width = \columnwidth]{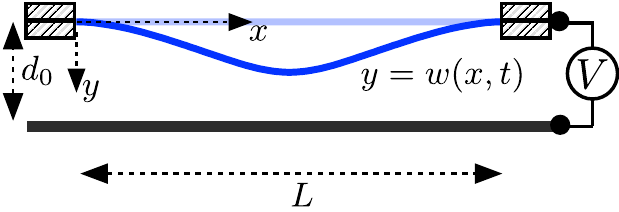} 
\caption{Schematic of a microbeam in its undeformed state (light blue) and deforming  under a DC load (dark blue). Here the ends of the beam are assumed to be clamped parallel to the lower electrode (shown as a thick black line).}
\label{fig:schematic}
\end{figure}

Microbeams are a widely used element of microelectromechanical systems (MEMS) \citep{pelesko2002}: they are a basic structural prototype that forms the building blocks for more complex devices \citep{lin2006}. In these applications they are subject to a range of loading types including magnetic, thermal and piezoelectric, though electrostatic forcing is the most commonly used \citep{das2009}. In a typical electrostatic device, the microbeam acts as a deformable electrode that is separated from a fixed electrode by a thin air gap (figure \ref{fig:schematic}). A potential difference is then applied between the electrodes. When the applied voltage exceeds a critical value, the microbeam collapses onto the fixed electrode during the `pull-in' instability \citep{batra2007}. Pull-in corresponds to a saddle-node (fold) bifurcation in which the stable shape away from collapse ceases to exist as an equilibrium solution \citep{zhang2014}. 

Microbeams are commonly used as microresonators in radio frequency (RF) applications, where a combination of AC and DC voltages drive the beam near its natural frequencies. In this context  pull-in generally corresponds to failure of the device \citep{nayfeh2007}. In switching applications, pull-in is instead exploited to generate large changes in shape between `off' and `on' states;  contact between the electrodes is prevented so pull-in can occur safely. Here it is important to understand the transient dynamics upon pull-in, as this governs the switching time of the device and hence the energy consumed during each cycle \citep{castaner1999}.

At voltages just beyond the pull-in voltage, a number of experiments and numerical simulations have reported that the transient dynamics slow down considerably compared to larger voltages \citep{gupta1996,gupta1997,gretillat1997,hung1999,younis2003,missoffe2008}. In this regime, the time taken to pull-in may increase by over an order of magnitude within a very narrow range of the applied voltage. While this is undesirable in switching applications, the slowing down is highly sensitive to ambient conditions, including air damping and the presence of external forces. This feature has been used to design an ambient pressure sensor based on measuring the pull-in time of a microbeam \citep{gupta1997}, and high-resolution accelerometers make use of similar behaviour in parallel-plate actuators \citep{rocha2004a,dias2011,dias2015}.

Despite the obvious potential as a sensing mechanism in MEMS, a detailed analysis of this slowing down has only been attempted for parallel-plate devices \citep{rocha2004a,rocha2004b}. Using a lumped mass-spring model, \cite{rocha2004a} identified a `metastable' or `bottleneck' phase that dominates the dynamics during pull-in: here the electrostatic force almost balances the mechanical restoring force, so the structure moves very slowly. In addition, this phase is found to only exist for devices that are overdamped (small quality factor), i.e.~when inertial effects are insignificant. However, many features of the bottleneck remain poorly understood --- for example it is not clear how the bottleneck duration scales with the applied voltage or other parameters of the system.

Using a lumped model similar to \cite{rocha2004a}, we showed in a previous study \cite{gomez2018} that the bottleneck phenomenon is caused by the remnant or `ghost' of the saddle-node bifurcation, similar to the `critical slowing down' observed in other physical systems such as elastic snap-through \citep{gomez2017a} and phase transitions \citep{chaikin}. Accordingly, the pull-in time, $t_{\mathrm{PI}}$, increases according to an inverse square-root scaling law \citep{strogatz}: we have $t_{\mathrm{PI}} \propto \epsilon^{-1/2}$ as $\epsilon \to 0$, where $\epsilon$ is the normalised difference between the applied voltage and the pull-in voltage. In addition, we determined an analytical expression for this pull-in time in terms of a lumped mechanical stiffness and effective damping coefficient appropriate to the bottleneck phase, which can then be used as fitting parameters to obtain good agreement with experiments and simulations of microbeams reported in the literature. However, this lumped-parameter approach does not show how the pull-in time depends on the various physical parameters of the beam (e.g.~its thickness and Young's modulus); such information would be useful when using the scaling law as a design rule in applications, as it eliminates the need for further simulations to predict the dynamic response if these parameters change. While it is possible to obtain equivalent stiffnesses under simple loading types (see \cite{larose2010}, for example), these do not account for effects such as a variable residual stress and different boundary conditions applied to the beam. An objective of this paper is thus to extend the lumped-parameter approach of \cite{gomez2018} to incorporate the beam geometry. 

Unlike lumped mass-spring models, it is much more difficult to make analytical progress with the equations governing microbeams; typically these consist of partial differential equations (PDEs) in space and time. A variety of numerical methods have therefore been developed to study the pull-in dynamics, including finite difference methods \citep{gupta1996,mccarthy2002}, finite element methods \citep{rochus2005} and reduced-order models (macromodels) \citep{nayfeh2005}. Macromodels typically apply a Galerkin procedure: the solution is expanded as a truncated series of known functions of the spatial variables (the basis functions), whose coefficients are unknown and depend on time. Commonly, the undamped vibrational modes of the undeformed beam are used as basis functions \citep{younis2003}. This yields a finite set of ordinary differential equations (ODEs) that can be integrated efficiently using pre-existing ODE solvers.

When only the first term in the Galerkin expansion is kept, this procedure results in a single-mode or single-degree-of-freedom (SDOF) approximation of the microbeam equations. Despite its simplicity, this approximation is often effective at capturing the leading-order dynamic phenomena --- for example the pull-in transition, the phase-plane portrait, and the influence of different parameter values and loading types have all been qualitatively explained using the SDOF method \citep{krylov2004,krylov2007,krylov2010}. Moreover, \cite{joglekar2011} have used the SDOF approximation to obtain an analytical expression for the pull-in time of an undamped microbeam. They found that using two different basis functions gives very similar results, suggesting that such approximations are reasonable. However, a comparison with numerical solutions indicated that the error in this approach grows larger near the pull-in transition, suggesting that a SDOF approximation is insufficient to model the dynamics near the pull-in transition. However, no systematic investigation of this error was provided in \cite{joglekar2011}. Elsewhere, the accuracy of the SDOF method has only been validated by computing natural frequencies and equilibrium shapes \citep{ijntema1992,kacem2009,batra2008}. It therefore remains unclear how valid the SDOF method is when analysing the transient dynamics of pull-in.

This motivates a more careful analysis of the pull-in dynamics of a microbeam. The key challenge we address in this paper is how to analyse the timescale of pull-in for a continuous elastic structure, without relying on detailed numerical simulations. We model the beam geometry using the dynamic beam equation, accounting for the effects of nonlinear midplane stretching and residual stress. However, similar to \cite{gomez2018}, we use a lumped damping coefficient to model the damping in the squeeze film between the beam and the lower electrode. While we could use a more complex damping model, this assumption enables us to make significant analytical progress. (The assumption of a constant damping coefficient can also be justified during the bottleneck phase, for reasons we shall discuss in \S\ref{sec:formulation}.) In particular, we develop an asymptotic method that reduces the governing PDE to a simpler ODE resembling the normal form for a saddle-node bifurcation. The key feature of this method is that the reduction is systematic and results in a SDOF-like approximation, but in which the appropriate basis function naturally emerges as part of the analysis. In light of this, we are then able to check the validity of SDOF approximations in which the basis function is chosen in an \emph{ad hoc} manner, as is standard in the literature. The asymptotic method also shows that the underlying bifurcation structure governs the bottleneck dynamics, rather than the precise physical details of the system, and so provides a general framework for analysing pull-in dynamics of microbeams and microplates in other loading scenarios.

The rest of this paper is organised as follows. We begin in \S\ref{sec:formulation} by describing the equations governing the microbeam dynamics and their non-dimensionalisation. In \S\ref{sec:equilibria}, we consider the equilibrium behaviour as the voltage is quasi-statically varied.  In \S\ref{sec:dynamics}, we analyse the dynamics when the voltage is just beyond the static pull-in voltage. Using direct numerical solutions, we demonstrate bottleneck behaviour in the overdamped limit. We then perform a detailed asymptotic analysis of the bottleneck phase. We confirm the expected scaling $t_{\mathrm{PI}} \propto \epsilon^{-1/2}$ as $\epsilon \to 0$, and we calculate the pre-factor in this relationship in terms of the beam parameters. This is compared to experiments and numerical simulations that incorporate more realistic models of squeeze film damping. In \S\ref{sec:singlemode} we consider the accuracy of a standard SDOF approximation. We demonstrate that the error of this approach can be small and we derive criteria that a `good' choice of basis function should satisfy.  Finally, we summarise our findings and conclude in \S\ref{sec:conclusions}.

\section{Theoretical formulation}
\label{sec:formulation}

\subsection{Governing equations}
A schematic of the microbeam is shown in figure \ref{fig:schematic}. The properties of the beam are its density $\rho_s$, thickness $h$, width $b$ and bending stiffness $B = E b h^3/12$, with $E$ the Young's modulus (using the bending stiffness appropriate for a narrow strip rather than an infinite plate \cite{audoly}). We suppose that the ends of the beam are clamped parallel to the lower electrode a distance $L$ apart (also called fixed--fixed ends \cite{gupta1997}). These boundary conditions are commonly used in applications of microbeams in pressure sensors and microswitches \citep{joglekar2011}, and have been widely studied as a `benchmark problem' \citep{krylov2008}. Because the natural length of the fabricated beam may differ slightly from $L$ \citep{lin2006}, we also account for a possible (constant) residual tension $P_0$ when the beam is flat ($P_0$ may also be negative, corresponding to residual compression). We choose coordinates so that $x$ measures the horizontal distance from the left end of the beam, and $y = w(x,t)$ is the transverse displacement (with $t$ denoting time). The applied DC voltage is $V$, and $d_0$ is the thickness of the air gap between the beam and the lower electrode in the absence of any displacement, $w = 0$ (figure \ref{fig:schematic}). 

When the microbeam passes through a bottleneck phase, the motions are dramatically slowed and so we can neglect compressibility and rarefaction effects in the squeeze film --- the damping is purely viscous \citep{missoffe2008}. As the geometry of the microbeam is also slowly varying in the bottleneck, we assume a constant damping coefficient, $\eta$. While it is possible to derive an approximate expression for the damping coefficient from the incompressible Reynolds equation \citep{blech1983,veijola1995,krylov2004}, we do not consider its precise form here, and instead treat $\eta$ as a lumped parameter for simplicity. This also allows us to parameterise additional effects such as material damping and different venting conditions at the beam edges. This approach is similar to that in \cite{gomez2018}, except we retain a complete description of the beam's shape here rather than using a lumped spring constant.

We assume the beam thickness is small compared to its length (i.e.~$h \ll L$) and its shape remains shallow; if the beam does not contact the lower electrode ($w < d_0$), this assumption is valid provided the aspect ratio of the air gap is also small, $d_0 \ll L$. Under the above assumptions, a vertical force balance on the beam yields the dynamic beam equation  \citep{pelesko2002}
\beq
\rho_s b h \pdd{w}{t} + \eta \pd{w}{t} + B\pdf{w}{x} - P \pdd{w}{x} = \frac{1}{2}\frac{\epsilon_0 b V^2}{(d_0-w)^2}, 
\label{eqn:beamdim}
\eeq
for $0 < x < L$, where $P(t)$ is the (unknown) tension in the beam and $\epsilon_0$ is the permittivity of air. Here we are using a parallel-plate approximation of the electrostatic force, consistent with our assumption  $d_0 \ll L$; for simplicity we neglect the effects of fringing fields (this requires $d_0 \ll b$) and we do not consider partial field screening between the beam and lower electrode. As the beam deforms, the tension associated with midplane stretching is (see e.g.~\citep{younis2003})
%
\beq
\frac{(P-P_0)L}{Ebh} = \frac{1}{2}\int_0^L \left(\pd{w}{x}\right)^2\:\id x,
\label{eqn:hookeslawdim}
\eeq
which we refer to as the Hooke's law constraint. The boundary conditions at the clamped ends are (subscripts denoting partial differentiation)
\beqn
w(0,t) = w_x(0,t) = w(L,t) = w_x(L,t) = 0. 
\eeqn
For initial conditions, we suppose the beam is at rest when the voltage is suddenly stepped from zero, i.e.~$w(x,0) = w_t(x,0) = 0$; these initial conditions are common in sensing applications \citep{gomez2018}. With these initial conditions, our model is equivalent to that of \cite{younis2003}, who focussed on solving the equations numerically using a reduced-order model. Here we will instead use the system to gain analytical understanding of the pull-in dynamics, including the bottleneck phenomenon. We will then use our own numerical solutions, together with those of \cite{younis2003}, to validate our results.

\subsection{Non-dimensionalisation}
It is convenient to scale the horizontal coordinate with the length $L$ between the clamps, and to scale the vertical displacement with the initial gap thickness $d_0$. As we are interested in overdamped devices, a natural timescale $[t]$ comes from balancing damping and bending forces in equation \eqref{eqn:beamdim}, giving $[t] = L^4 \eta /B$.   We therefore introduce the dimensionless variables 
\beqn
X = \frac{x}{L}, \quad W = \frac{w}{d_0}, \quad T = \frac{t}{[t]}.
\eeqn
With these re-scalings, the beam equation \eqref{eqn:beamdim} becomes
\beq
Q^2 \pdd{W}{T} + \pd{W}{T} + \pdf{W}{X} - \tau \pdd{W}{X} = \frac{\lambda}{(1-W)^2}, 
\label{eqn:beam}
\eeq
for $0 < X < 1$, where we introduce the dimensionless parameters 
\beq
Q = \frac{\sqrt{\rho_s b h B/L^4}}{\eta}, \quad \tau(T) = \frac{P L^2}{B}, \quad \lambda = \frac{1}{2}\frac{\epsilon_0 b L^4 V^2}{B d_0^3}.  
\label{eqn:parameters} 
\eeq
These correspond to the quality factor, the dimensionless tension in the beam, and the dimensionless voltage, respectively. We may interpret $\lambda$ as the ratio of the typical electrostatic force per unit length ($\sim \epsilon_0 b V^2 /[2 d_0^2]$) to the typical force per unit length required to bend the beam by an amount comparable to $d_0$ ($\sim B d_0 / L^4$).  


Re-scaling the Hooke's law constraint \eqref{eqn:hookeslawdim}, the dimensionless tension $\tau$ is given by
\beq
\cS (\tau - \tau_0) = \frac{1}{2}\int_0^1 \left(\pd{W}{X}\right)^2\:\id X, \label{eqn:hookeslaw}
\eeq
where
\beqn
\tau_0 = \frac{P_0 L^2}{B}, \quad \cS  = \frac{h^2}{12 d_0^2},
\eeqn
are the dimensionless residual tension and `stretchability' of the beam \citep{pandey2014}. Here $\mathcal{S}$ acts as a dimensionless membrane stiffness. In  real devices the beam thickness $h$ is often comparable to the initial gap thickness $d_0$ \citep{gupta1996,gupta1997}, so that $\mathcal{S}$ typically lies in the range $(10^{-2},10^{-1})$. Finally, the boundary conditions  at the clamped ends and initial conditions become 
\begin{eqnarray}
W(0,T) = W_X(0,T) = W(1,T) = W_X(1,T) = 0, \label{eqn:bc} \\
W(X,0) = W_T(X,0) = 0.  \label{eqn:ic}
\end{eqnarray}

\section{Equilibrium behaviour}
\label{sec:equilibria}
We briefly review the equilibrium behaviour as the dimensionless voltage $\lambda$ is quasi-statically varied. We solve the steady version of the beam equation \eqref{eqn:beam}
together with the Hooke's law constraint \eqref{eqn:hookeslaw} and boundary conditions \eqref{eqn:bc} numerically in \textsc{matlab} using the routine \texttt{bvp4c}. We write the beam equation as a first-order system in $W$ and its derivatives, and we impose \eqref{eqn:hookeslaw}  by introducing the additional variable $I'(X) = \frac{1}{2} \left[W'(X)\right]^2$ (writing $'$ for $\mathrm{d}/\mathrm{d}X$) with boundary conditions $I(0) = 0$ and $I(1) = \cS(\tau-\tau_0)$.  Because pull-in corresponds to a saddle-node bifurcation, near which the system becomes highly sensitive to $\lambda$, we avoid convergence issues \citep{younis2003} by instead controlling the tension $\tau$ and solving for $\lambda$ as part of the solution (such unknown parameters are easily incorporated into the \texttt{bvp4c} solver). For each stretchability $\cS$ and residual tension $\tau_0$, we implement a simple continuation algorithm that follows equilibrium branches as $\tau$ is increased in small steps. For an initial guess to begin the continuation, we use an asymptotic solution valid at small voltages, when the beam is nearly flat and $\tau \approx \tau_0$.

\begin{figure*}
\centering
\includegraphics[width = \textwidth]{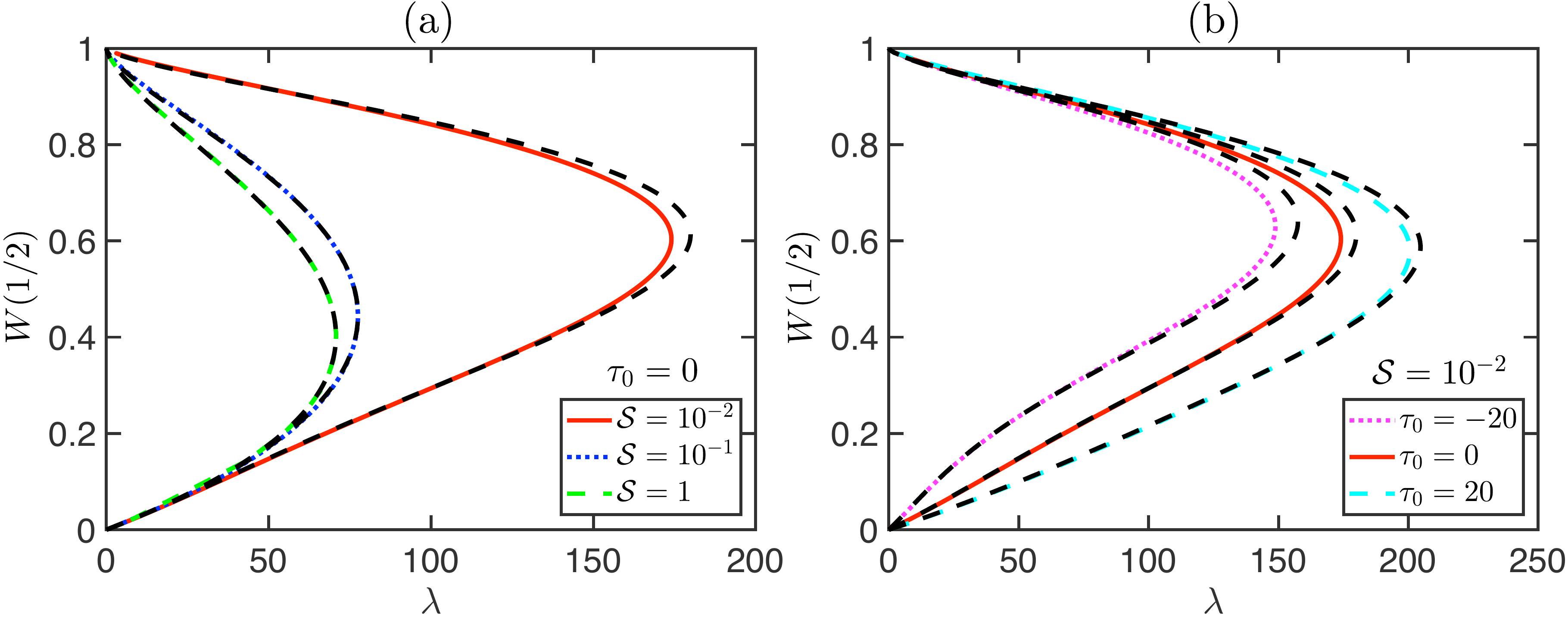} 
\caption{Response diagram for steady solutions of the beam equation \eqref{eqn:beam} subject to \eqref{eqn:hookeslaw}--\eqref{eqn:bc} as the dimensionless voltage $\lambda$ varies. Numerical results are shown for (a) zero residual tension ($\tau_0 = 0$) and varying stretchability $\cS$ and (b) fixed stretchability $\cS = 10^{-2}$ and varying residual tension $\tau_0$ (coloured curves as in legends). For later comparison, predictions from the SDOF approximation computed using equation \eqref{SDOFsteadysoln} are shown (black dashed curves).}
\label{fig:response}
\end{figure*}

When plotted back in terms of $\lambda$, the resulting bifurcation diagram confirms that for small $\lambda$, two distinct, physical (i.e.~$W < 1$) equilibrium branches exist. As $\lambda$ increases, both branches approach each other, before they eventually meet at a saddle-node bifurcation when $\lambda = \lamf$: no equilibrium shape away from collapse exists for $\lambda  > \lamf$ (we are unable to numerically find further solutions). This is shown in figure \ref{fig:response}a, where we plot the midpoint displacement, $W(1/2)$, as a function of $\lambda$. The critical value $\lamf$ evidently increases as $\cS$ decreases (corresponding to a larger membrane stiffness), growing rapidly for values $\cS \lesssim 10^{-1}$. At the small stretchabilities typical of realistic devices, the dependence of $\lamf$ on the residual tension $\tau_0$ is much weaker; see figure \ref{fig:response}b. (For later reference, in both plots we also show the predictions of the SDOF approximation computed in \S\ref{sec:singlemode}.)

Using a standard linear stability analysis, it has been shown \citep{younis2003} that the equilibrium branches below the fold point in figures \ref{fig:response}a--b (i.e.~with $W(1/2)\to 0 $ as $\lambda\to 0$) are linearly stable and correspond to the shapes observed experimentally. The upper branches are linearly unstable, so the fold point corresponds to a standard `exchange of stability' \citep{maddocks1987} in which both branches become neutrally stable as they meet. For later reference, we note that only the fundamental natural frequency (eigenvalue) of the beam equals zero at the fold, so the zero eigenvalue there is simple (i.e. the eigenspace is of dimension one). We deduce that the critical value $\lamf$  corresponds to where pull-in \emph{first} occurs if $\lambda$ is increased quasi-statically. In dimensional terms, this gives the static pull-in voltage, $V_{\mathrm{SPI}}$, and the pull-in displacement, $ w_{\mathrm{SPI}}(x)$, as 
\beqn
V_{\mathrm{SPI}} = \sqrt{\frac{2 B d_0^3 \lamf}{\epsilon_0 b L^4}}, \quad w_{\mathrm{SPI}}(x) = d_0 \Wf (X),
\eeqn
where we write $\Wf(X)$ for the dimensionless equilibrium shape at the fold point (with associated tension $\tauf$).

\section{Pull-in dynamics}
\label{sec:dynamics}
We now explore the dynamics at voltages just beyond the static pull-in transition, setting
\beqn
\lambda =\lamf(1+\epsilon),
\eeqn
where $0 < \epsilon \ll 1$ is a small perturbation. If all parameters except the voltage are fixed, combining the definition of $\lambda$ in \eqref{eqn:parameters} with the fact that $\lambda_{\mathrm{fold}} = \epsilon_0 b L^4 V_{\mathrm{SPI}}^2/(2B d_0^3)$
shows that $\epsilon$ is simply the normalised voltage difference:
\beqn
\epsilon = \frac{\lambda}{\lambda_{\mathrm{fold}}}-1 = \left(\frac{V}{V_{\mathrm{SPI}}}\right)^2-1 \approx \frac{2}{V_\mathrm{SPI}} \left(V-V_\mathrm{SPI}\right).
\eeqn

We solve the dynamic beam equation \eqref{eqn:beam} subject to \eqref{eqn:hookeslaw}--\eqref{eqn:ic} numerically using the method of lines \cite{morton}. This involves discretising the equations using finite differences in space, so that the system reduces to a finite set of ODEs in time. We obtain second-order accuracy in the convergence of our scheme; for details see Appendix A. For each combination of $Q$, $\lambda$, $\mathcal{S}$ and $\tau_0$, we integrate the ODEs numerically in \textsc{matlab} (routine \texttt{ode23t}) to compute the trajectory of each grid point in the discretisation. To avoid the singularity at $W = 1$, we use event location to stop integration as soon as $(1-W) < \mathrm{tol}$ at any grid point, for some specified tolerance $\mathrm{tol}$. The corresponding time at this event is then the reported pull-in time, labelled $T_{\mathrm{PI}}$. For all simulations reported in this paper we use $N = 100$ grid points and $\mathrm{tol} = 10^{-4}$; we also specify relative and absolute error tolerances of $10^{-8}$ in \texttt{ode23t} and we limit the maximum dimensionless time step of the solver to $10^{-6}$. We have checked that our results are insensitive to further increasing $N$ and decreasing these tolerances/maximum time step. 

\begin{figure*}
\centering
\includegraphics[width = \textwidth]{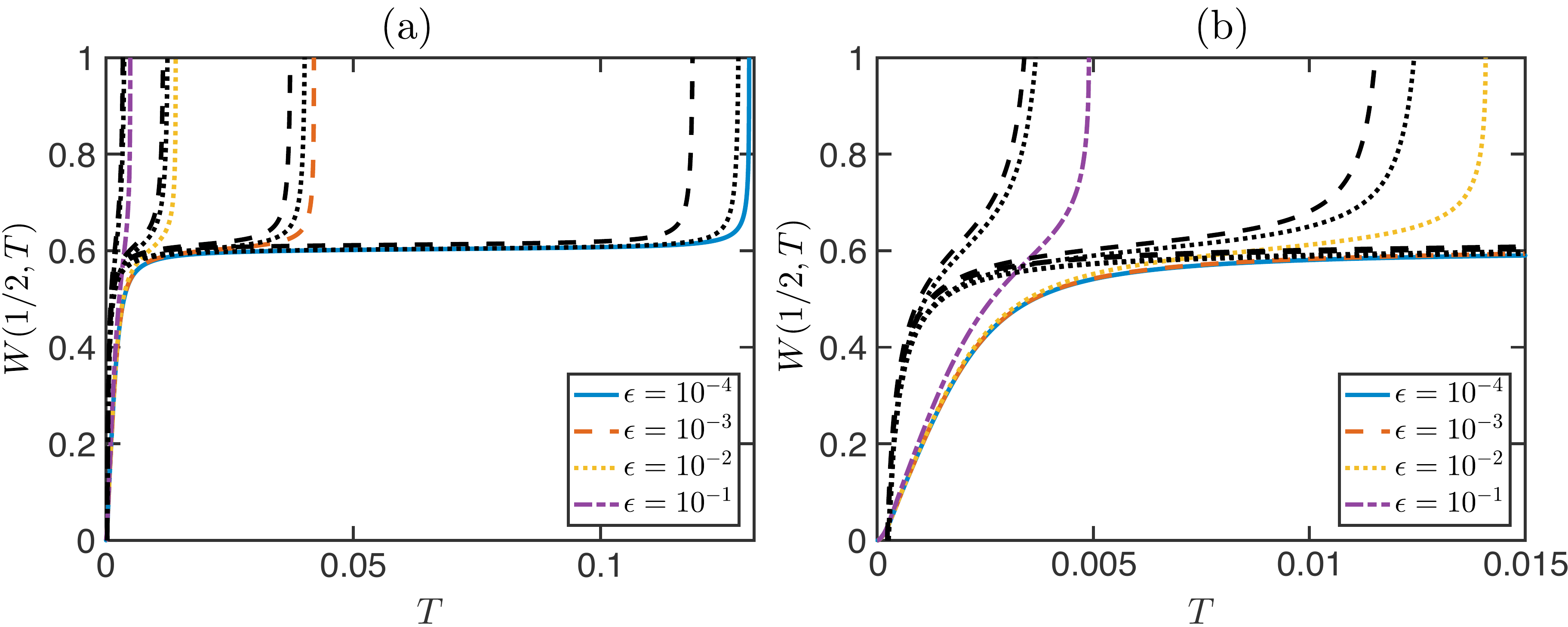} 
\caption{Bottleneck behaviour at voltages close to the pull-in transition ($Q = 10^{-2}$, $\cS = 10^{-2}$, $\tau_0 = 0$, $N = 100$). (a) Dimensionless midpoint trajectories obtained by integrating the dynamic beam equation \eqref{eqn:beam} subject to \eqref{eqn:hookeslaw}--\eqref{eqn:ic} numerically (coloured curves as in legends). These exhibit a bottleneck as $W(1/2,T)$ passes $\Wf(1/2)\approx 0.6036$, which increases in duration as $\epsilon$ decreases. For later comparison, also shown are the predictions \eqref{eqn:bottleneckdisplacement}  of the bottleneck analysis (black dotted curves) and the predictions \eqref{eqn:SDOFdisplacement} of the SDOF approximation (black dashed curves). (b) A close up of the trajectories in (a) at early times.}
\label{fig:trajectories}
\end{figure*}

\begin{figure}
\centering
\includegraphics[width = \columnwidth]{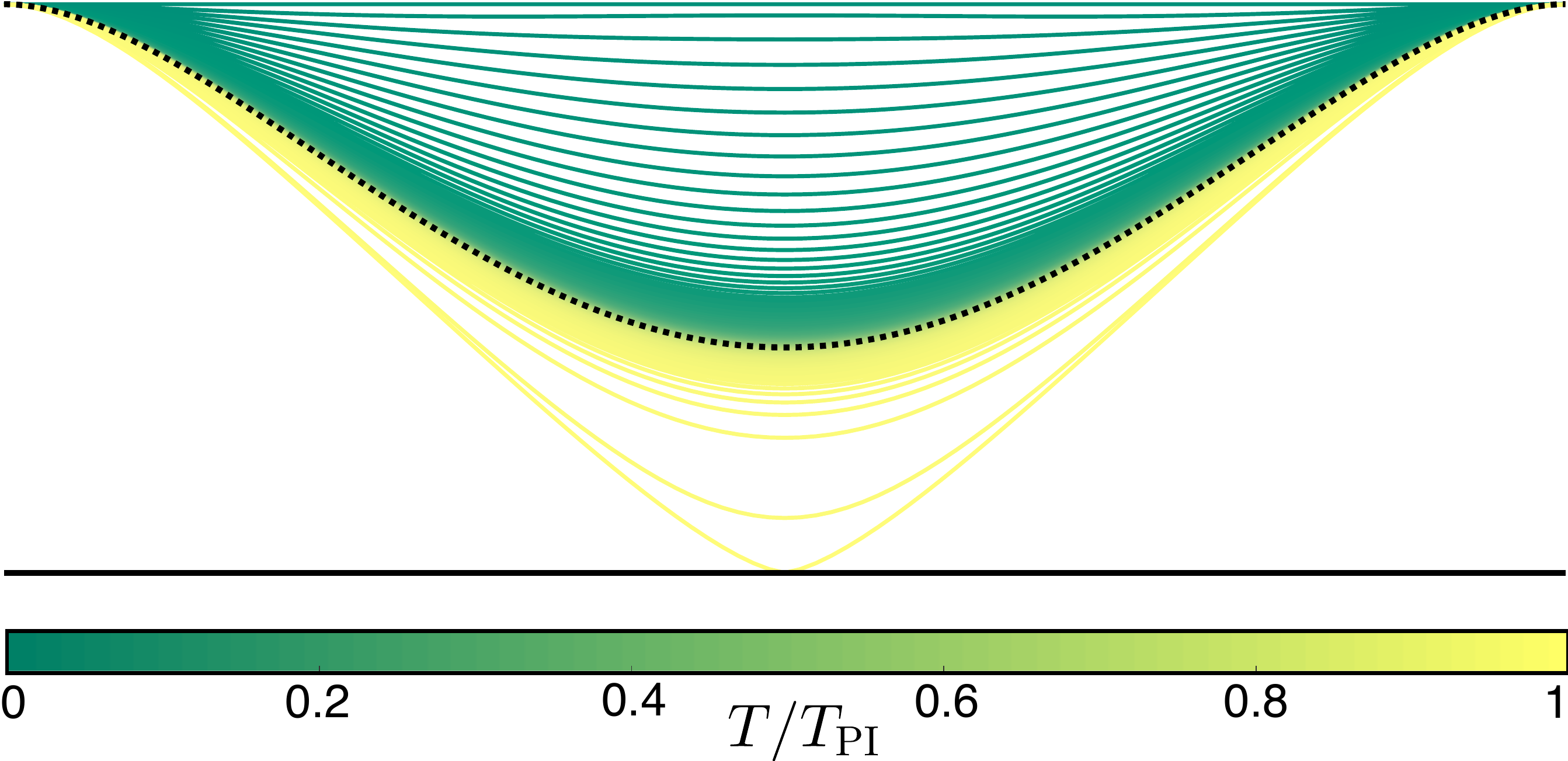} 
\caption{Sequence of numerically-determined beam shapes during pull-in ($\epsilon = 10^{-3}$, $Q = 10^{-2}$, $\cS = 10^{-2}$, $\tau_0 = 0$, $N = 100$). In total, $212$ profiles at equally spaced time steps between $T = 0$ and contact with the lower electrode (shown as a black line) at $T = T_{\mathrm{PI}}$ are displayed (coloured curves; see colourbar), as well as the pull-in displacement $\Wf(X)$ (black dotted curve).}
\label{fig:shapes}
\end{figure}

Numerical trajectories of the beam midpoint, $W(1/2,T)$, are plotted in figure \ref{fig:trajectories} for various values of $\epsilon$ (here we have set $Q = 10^{-2}$, corresponding to an overdamped beam). We observe that the microbeam slows down significantly in a bottleneck phase. This is similar to the bottleneck behaviour of a parallel-plate capacitor studied in \cite{gomez2018}, in that (i) the bottleneck dominates the total time taken to pull-in; (ii) the duration of the bottleneck is highly sensitive to the value of $\epsilon$, increasing apparently without bound as $\epsilon \to 0$; and (iii) the bottleneck always seems to occur close to a well-defined displacement. Indeed, this displacement is precisely the static pull-in displacement; see figure \ref{fig:shapes}, which shows that the beam slows down dramatically near the fold shape, $W_{\mathrm{fold}}(X)$ (black dotted curve), before rapidly accelerating towards the lower electrode, as seen by the shapes (plotted at equally spaced times) becoming closely packed together. 

These similarities suggest that the bottleneck here is also a saddle-node ghost: as the beam passes the static pull-in displacement, the net force becomes very small (since $\epsilon \ll 1$ and the forces balance exactly at the pull-in displacement with $\epsilon = 0$), so that the motions slow down considerably. Hence, we expect that inertia of the beam does not play a role. We now perform a detailed analysis of the solution during the bottleneck phase. We use a similar method to \cite{gomez2018}: we expand the solution about the pull-in displacement, and solve the governing equations asymptotically. However, the system here is infinite dimensional and the pull-in displacement is the function $W_{\mathrm{fold}}(X)$ rather than a lumped scalar value. It turns out that the `extra' degrees of freedom mean we need to proceed to higher order to obtain a simple equation that characterises the bottleneck dynamics. This will allow us not only to obtain the expected $\epsilon^{-1/2}$ scaling for the bottleneck duration, but also to calculate the corresponding pre-factor.

\subsection{Bottleneck analysis}
\label{sec:bottleneckanalysis}
When the solution is close to the static pull-in displacement, we have
\begin{eqnarray}
W(X,T) &  = & \Wf(X) + \Wt(X,T),  \nonumber \\
\tau(T) & = & \tauf + \taut(T),  \label{eqn:bottleneckexpansions}
\end{eqnarray}
where $|\Wt| \ll 1$ and $|\taut| \ll 1$.  It follows that the electrostatic force can be expanded as
\begin{eqnarray}
\frac{\lambda}{(1-W)^2} & = & \frac{\lamf}{(1-\Wf)^2}\left(1+\epsilon\right) + \frac{2 \lamf}{(1-\Wf)^3}\Wt \nonumber\\ 
&& \: + \frac{3\lamf}{(1-\Wf)^4}\Wt^2 + O(\epsilon \Wt,\Wt^3). \label{eqn:bottleneckexpandforce}
\end{eqnarray}
(The reason why we retain the $O(\Wt^2)$ term but neglect the $O(\epsilon \Wt,\Wt^3)$ terms will be discussed below.) Inserting these expansions into the dynamic beam equation \eqref{eqn:beam}, and neglecting the inertia term (which from the above discussion is not expected to be important in the bottleneck), we obtain
\begin{eqnarray}
L(\Wt,\taut) &= & -\pd{\Wt}{T}+\taut \pdd{\Wt}{X} + \frac{\lamf}{(1-\Wf)^2}\epsilon \nonumber \\
&& \: +\frac{3\lamf}{(1-\Wf)^4}\Wt^2 + O(\epsilon \Wt,\Wt^3), \label{eqn:bottleneckbeam}
\end{eqnarray}
where we have introduced the linear operator
\beq
L(U,V) \equiv \pdf{U}{X}-\tauf \pdd{U}{X} - V\dd{\Wf}{X} - \frac{2 \lamf U}{(1-\Wf)^3}.
\label{eqn:bottleneckdefnL}
\eeq
The Hooke's law constraint \eqref{eqn:hookeslaw} becomes
\beq
\mathcal{S}\taut = \int_0^1 \d{\Wf}{X}\pd{\Wt}{X}~\id X + \frac{1}{2}\int_0^1 \left(\pd{\Wt}{X}\right)^2~\id X, \label{eqn:bottleneckhookeslaw}
\eeq
and the boundary conditions \eqref{eqn:bc} imply that 
\beq
\Wt(0,T) = \Wt_X(0,T) = \Wt(1,T) = \Wt_X(1,T) = 0. \label{eqn:bottleneckbc}
\eeq

We now make two important assumptions that we will check at the end of our analysis: 
\begin{enumerate}[label=(\roman*)]
\item{For small perturbations $\epsilon \ll 1$ the bottleneck timescale satisfies $T \gg 1$.}
\item{In the bottleneck, we must account for changes in the solution that are much larger than $\epsilon$ but remain small compared to unity, i.e.~$\epsilon \ll |\Wt| \ll 1$ and $\epsilon \ll |\taut| \ll 1$.}
\end{enumerate}
These assumptions are partly justified by the analysis in \citep{gomez2018}, which showed that (i) and (ii) hold for a parallel-plate capacitor. The idea is that while $\Wt$ is $O(\epsilon)$ on smaller, inner, timescales, these assumptions will allow us to correctly predict the total bottleneck duration, when we later compare the results to numerics.
In particular, these assumptions imply that the right-hand side of equation \eqref{eqn:bottleneckbeam} remains small: the time derivative is small by virtue of the slow bottleneck timescale, while the remaining terms are either quadratic in the small quantities $(\Wt,\taut)$, or are $O(\epsilon)$. The left-hand side is linear in $(\Wt,\taut)$ and hence dominates these terms (from assumption (ii)). We now use this property to solve the problem asymptotically. 

\subsubsection{Leading order:}
We expand
\begin{eqnarray}
\Wt(X,T) &  \sim & \Wt_0(X,T) + \Wt_1(X,T), \nonumber \\
\taut(T) & \sim & \taut_0(T) + \taut_1(T),  \label{expandWttaut}
\end{eqnarray}
where $|\Wt_1| \ll |\Wt_0|$ and $|\taut_1| \ll |\taut_0|$ are first-order corrections. From the above discussion, at leading order we then have the homogeneous problem 
\beqn
L(\Wt_0,\taut_0) = 0.
\eeqn
The constraint \eqref{eqn:bottleneckhookeslaw} at leading order is
\beqn
\cS \taut_0 = \int_0^1 \d{\Wf}{X}\pd{\Wt_0}{X}~\id X,
\eeqn
while the clamped conditions \eqref{eqn:bottleneckbc} remain unchanged in terms of $\Wt_0$.  These leading order equations are precisely the homogeneous, linearised versions of the full system \eqref{eqn:bottleneckbeam}--\eqref{eqn:bottleneckbc} in $(\Wt,\taut)$. Hence, they are equivalent to the equations governing linear stability of the fold shape $(\Wf,\tauf)$, but --- crucially --- restricted to neutrally-stable modes (eigenfunctions) whose natural frequency (eigenvalue) is zero: we would have obtained similar equations for $(W_p,\tau_p)$ upon setting $W = \Wf(X)+\delta W_p(X)e^{i\omega T}$ and $\tau = \tauf + \delta \tau_p e^{i\omega T}$ in the original beam equations, considering terms of $O(\delta)$ and setting $\omega = 0$. 

Recall from our earlier discussion in \S\ref{sec:equilibria} that only the fundamental natural frequency equals zero at the fold bifurcation. The homogeneous problem in $L(\cdot,\cdot)$ therefore has a one-dimensional solution space, spanned by the pair $(W_p,\tau_p)$ satisfying
\begin{eqnarray}
&& L(W_p,\tau_p) = 0,~W_p(0) = W_p'(0) = W_p(1) = W_p'(1) = 0,  \nonumber \\
&& \cS \tau_p = \int_0^1 \d{\Wf}{X}\d{W_p}{X}~\id X, \quad \int_0^1 W_p^2~\id X  = 1. \label{eqn:neutralstability}
\end{eqnarray}
(The final equation here is a normalisation condition required to uniquely specify $W_p$). While this seems to over-determine the eigenfunction $W_p$ ($L(\cdot,\cdot)$  is fourth order and $\tau_p$ is unknown, but we have six constraints), we are guaranteed a solution when $\Wf$ is specifically the equilibrium shape evaluated at the fold. We deduce that the solution for $(\Wt_0,\taut_0)$ must be a multiple of the pair  $(W_p,\tau_p)$:
\beqn
(\Wt_0,\taut_0) = A(T) (W_p,\tau_p),
\eeqn
for some variable $A(T)$. The variable $A(T)$ plays a key role in the pull-in dynamics: re-arranging the original series expansion in \eqref{eqn:bottleneckexpansions} shows that
\beq 
A(T) = \frac{\Wt_0(X,T)}{W_p(X)} \sim \frac{W(X,T)-\Wf(X)}{W_p(X)}, \label{eqn:bottleneckdefnA}
\eeq
so that $A(T)$ characterises how the beam evolves away from the pull-in displacement during the bottleneck. Equation \eqref{eqn:bottleneckdefnA} also shows how we have performed a SDOF-type approximation: the solution is projected onto the neutrally-stable eigenfunction $W_p$ associated with the loss of stability at the fold. 
Currently we have not yet determined the amplitude $A(T)$. As with other problems in elasticity, such as Euler buckling of a straight beam \citep{howell}, we expect to determine  $A(T)$ using a solvability condition on a higher order problem.

\subsubsection{First order:}
To obtain the first-order problem, we substitute the expansions \eqref{expandWttaut} into equation \eqref{eqn:bottleneckbeam} and neglect higher-order terms in favour of those involving the leading-order terms $(\Wt_0,\taut_0)$. The result is the same operator $L(\cdot,\cdot)$ as in the leading-order problem, though now applied to  $(\Wt_1,\taut_1)$, together with an inhomogeneous right-hand side forced by the leading-order terms. 
To obtain non-trivial dynamics at leading order, i.e.~for which $A(T) \neq \mathrm{constant}$, it is necessary to include both the time derivative and $O(\epsilon)$ terms at this order. Substituting $(\Wt_0,\taut_0) = A(T) (W_p,\tau_p)$ gives
\begin{eqnarray}
L(\Wt_1,\taut_1) & = & -W_p\d{A}{T} + \frac{\lamf}{(1-\Wf)^2}\epsilon \nonumber \\
&& \: + \left[\tau_p\dd{W_p}{X} + \frac{3\lamf}{(1-\Wf)^4}W_p^2\right]A^2.  \label{eqn:bottleneckbeamfirst}
\end{eqnarray}
(Assumption (ii) above guarantees that the neglected terms of $O(\epsilon\Wt_0,\Wt_0^3)$ in \eqref{eqn:bottleneckexpandforce} are small compared to the terms retained here.) Similarly, the Hooke's law constraint \eqref{eqn:bottleneckhookeslaw} at first order can be written as
\beq
\cS \taut_1  - \int_0^1 \d{\Wf}{X}\pd{\Wt_1}{X}~\id X = \frac{A^2}{2}\int_0^1 \left(\d{W_p}{X}\right)^2~\id X, \label{eqn:bottleneckhookeslawfirst}
\eeq
while the clamped boundary conditions \eqref{eqn:bottleneckbc} remain unchanged in terms of $\Wt_1$.

The first-order problem is of the form $L\mathbf{y} = f$, where $\mathbf{y} \equiv (\Wt_1,\taut_1)$, with linear boundary conditions/constraints in the components of $\mathbf{y}$. Because the homogeneous problem $L\mathbf{y} = 0$ has the non-trivial solution $(W_p,\tau_p)$, the Fredholm Alternative Theorem \citep{keener} states that solutions to the inhomogeneous problem  can only exist for a certain function $f$. This yields a solvability condition that takes the form of an ODE for $A(T)$. We formulate this condition in the usual way: we multiply \eqref{eqn:bottleneckbeamfirst} by a solution of the homogeneous adjoint problem (it may be verified that  $L(\cdot,\cdot)$ is self-adjoint, so one solution is simply  $W_p$), integrate over the domain, and use integration by parts to shift the operator onto the adjoint solution.  
%
Upon simplifying, using  $L(W_p,\tau_p) = 0$, the  clamped boundary conditions and Hooke's law constraints satisfied by $\Wt_1$, $W_p$, $\Wf$, and the normalisation $\int_0^1 W_p^2 ~\id X = 1$, we arrive at
\beq
\d{A}{T} = c_1 \epsilon + c_2 A^2, \label{eqn:bottleneckode}
\eeq
where 
\begin{eqnarray}
c_1 & = & \lamf \int_0^1 \frac{W_p}{(1-\Wf)^2}~\id X,  \nonumber \\
c_2 & = & 3\lamf  \int_0^1 \frac{W_p^3}{(1-\Wf)^4}~\id X  \nonumber \\
&& \: - \frac{3}{2\mathcal{S}}\left[\int_0^1\d{\Wf}{X} \d{W_p}{X}\id X\right]\left[\int_0^1 \left(\d{W_p}{X}\right)^2\id X\right]. \nonumber \\
\label{eqn:defnc1c2}
\end{eqnarray}

We have therefore reduced the leading-order dynamics in the bottleneck to the normal form for a saddle-node bifurcation (up to numerical constants) \citep{strogatz}, resembling a single family of ODEs parameterised by $\epsilon$. Note that if we had not included the time derivative and the  $O(\epsilon)$ term in \eqref{eqn:bottleneckbeamfirst}, but left these to a higher-order problem, we would have obtained trivial dynamics at this stage with \eqref{eqn:bottleneckode} instead giving $A  = 0$. We also note that $\epsilon$ can be scaled out of the normal form by setting $A = \epsilon^{1/2}\mathcal{A}$ and  $T = \epsilon^{-1/2}\mathcal{T}$. Retracing our steps above, this implies that the leading-order and first-order problems are obtained at $O(\epsilon^{1/2})$ and $O(\epsilon)$ respectively. We could have obtained the same equations by simply posing a regular expansion of the solution in powers of $\epsilon^{1/2}$. This is essentially the approach used to analyse bottleneck dynamics in other physical systems \citep{aranson2000,gomez2017a}; our analysis here explains why this is the correct expansion sequence to use. 

The normal form \eqref{eqn:bottleneckode} also resembles the equation derived by \cite{gomez2018} for a lumped mass-spring model, and by \citep{krylov2004,zaitsev2012} in other pull-in problems. This provides further evidence that \eqref{eqn:bottleneckode} is generic  for the dynamics of pull-in in overdamped devices. We see that the precise form of the boundary conditions applied to the microbeam enters only through the constants $c_1$ and $c_2$ (as the boundary conditions determine the eigenfunction $W_p$ and fold shape $\Wf$). For each stretchability $\cS$ and  residual tension $\tau_0$, we evaluate these by solving the neutral stability problem \eqref{eqn:neutralstability} numerically (using \texttt{bvp4c}) and using quadrature to evaluate the integrals appearing in $c_1$ and $c_2$.

\subsubsection{Solution for $A(T)$:}
The solution of \eqref{eqn:bottleneckode} is
\beq
A = \sqrt{\frac{c_1 \epsilon}{c_2}} \tan\Big[\sqrt{c_1 c_2 \epsilon}(T-T_0)\Big], \label{eqn:bottlenecksolnA}
\eeq
for some constant $T_0$. At this stage, we can check when our original assumption (ii) holds, i.e.~when the leading-order solution satisfies $\epsilon \ll |\Wt_0| \ll 1$ and $\epsilon \ll |\taut_0| \ll 1$. From the expression $\Wt \sim \Wt_0 = W_p(X) A(T)$, this requires $\epsilon \ll |A| \ll 1$ (since $W_p$ is $O(1)$). However, further analysis (given in Appendix B) shows that the solution \eqref{eqn:bottlenecksolnA} also applies when $A = O(\epsilon)$. We therefore only require $|A| \ll 1$, i.e.~
\beqn
\left| \tan\Big[\sqrt{c_1 c_2 \epsilon}(T-T_0)\Big]\right | \ll \epsilon^{-1/2}.
\eeqn
This breaks down when the tan function is very large; the expansion  $\tan x \sim \pm (\pi/2\mp x)^{-1}$ as $x \to \pm\pi/2$ implies that this occurs when
\beqn
T - T_0 \sim \pm \frac{\pi}{2\sqrt{c_1 c_2 \epsilon}}. \label{eqn:solnAvalidity}
\eeqn
At this point, the amplitude $A$ reaches $O(1)$ and our asymptotic analysis breaks down. Because $A$ is growing rapidly by this stage (according to the tan function), the beam is no longer in the bottleneck phase. The minus sign here therefore corresponds to initially entering the bottleneck, while the plus sign corresponds to exiting the bottleneck towards pull-in. The duration of the bottleneck is thus
\beqn
T_{\mathrm{bot}} \sim \frac{\pi}{\sqrt{c_1 c_2 \epsilon}}.
\eeqn
(This validates our earlier assumption (i) that the bottleneck timescale satisfies $T \gg 1$ when $\epsilon \ll 1$.) The trajectories shown in figure \ref{fig:trajectories} suggest that the bottleneck dominates all other timescales in the problem; this includes transients around $T = 0$ and just before contact where inertia is important. The dimensionless pull-in time, $T_{\mathrm{PI}}$, to leading order is then the bottleneck duration,
\beq
T_{\mathrm{PI}} \sim \frac{\pi}{\sqrt{c_1 c_2 \epsilon}}.  \label{eqn:bottlenecktime}
\eeq
Because the solution for $A$ is antisymmetric about $T_0$, it also follows that $T_0$ is simply half of the bottleneck duration: $T_0 \sim \pi/(2\sqrt{c_1 c_2 \epsilon})$. The solution \eqref{eqn:bottlenecksolnA} can then be written as 
\beqn
A \sim \sqrt{\frac{c_1 \epsilon}{c_2}} \tan\Big[\sqrt{c_1 c_2 \epsilon}~T-\frac{\pi}{2}\Big].
\eeqn
Writing this back in terms of the dimensionless displacement $W$ (see \eqref{eqn:bottleneckdefnA}), we therefore have
\beq 
W(X,T) \sim \Wf(X) + \sqrt{\frac{c_1 \epsilon}{c_2}} W_p(X) \tan\Big[\sqrt{c_1 c_2 \epsilon}~T-\frac{\pi}{2}\Big]. \label{eqn:bottleneckdisplacement}
\eeq

\subsection{Comparison with numerical results}
To compare our predictions to direct numerical solutions, we consider the case $\cS = 10^{-2}$ and  zero residual tension, $\tau_0 = 0$. We compute
\begin{eqnarray}
\lamf \approx 174.0343 , \quad \Wf(1/2) \approx 0.6036, \nonumber \\
W_p(1/2) \approx 1.571, \quad c_1 \approx 601.2, \quad c_2 \approx 9985. \label{eqn:S1e-2tau0values}
\end{eqnarray}
Using these values, for a specified $\epsilon$ we determine the midpoint displacement in the bottleneck using \eqref{eqn:bottleneckdisplacement}. The predicted behaviour is superimposed (as black dotted curves) onto numerical trajectories in figures \ref{fig:trajectories}a--b. We see that for $\epsilon \lesssim 10^{-3} $ the agreement is excellent during the bottleneck phase, i.e.~while $W(1/2,T)$ remains close to $\Wf(1/2) \approx 0.6036$; outside of this interval, the agreement breaks down as the bottleneck analysis is no longer asymptotically valid. In particular, very close to $T = 0$ and $T = T_{\mathrm{PI}}$, the asymptotic predictions become unbounded and diverge from the numerics.

\begin{figure*}
\centering
\includegraphics[width = \textwidth]{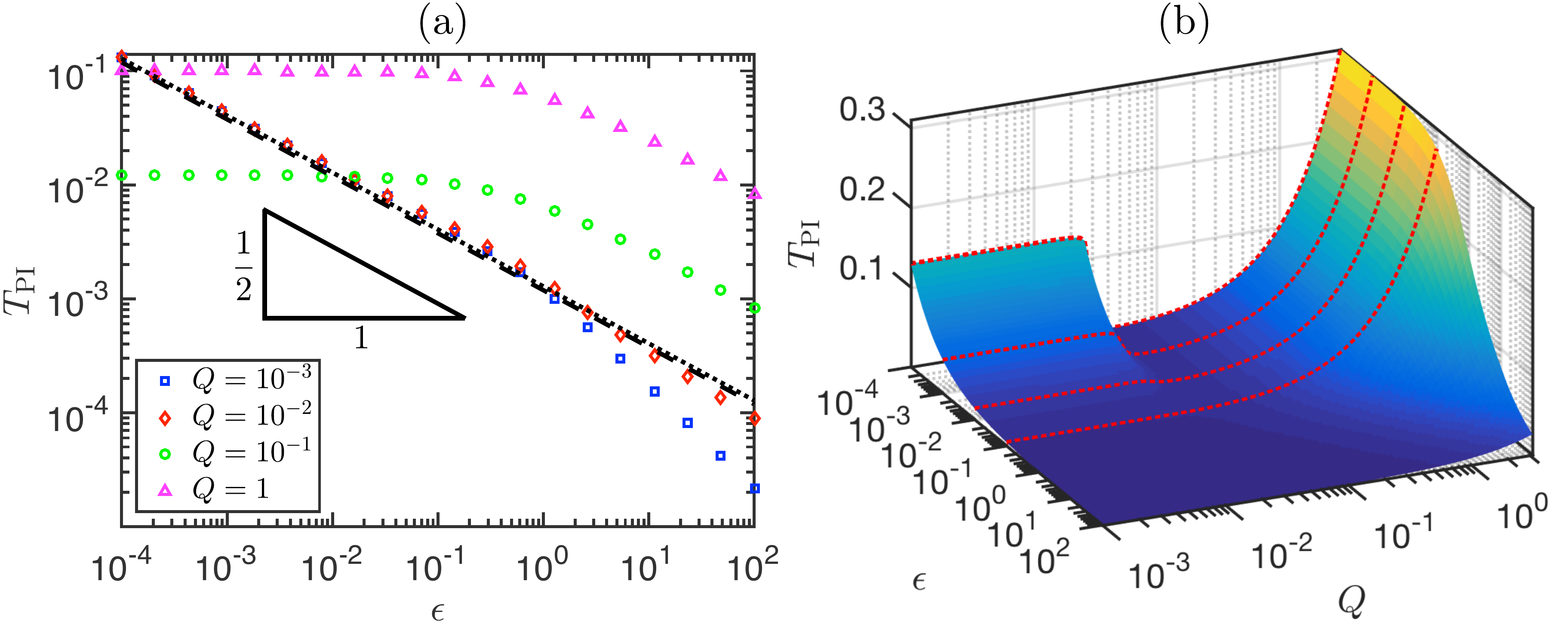} 
\caption{Pull-in times at voltages close to the pull-in transition ($\cS = 10^{-2}$, $\tau_0 = 0$, $N = 100$). (a) Numerical results for fixed $Q$ and variable $\epsilon$ (symbols as in legend). Also shown is the asymptotic prediction \eqref{eqn:bottlenecktime} from the bottleneck analysis (black dotted line), and, for later comparison, the prediction \eqref{eqn:SDOFpullintime} from the SDOF approximation (black dashed line) which is almost indistinguishable. (b) Surface plot of the numerical pull-in times. Slices through the surface (red dotted curves) are shown at $\epsilon \in \left\lbrace 10^{-4},10^{-3},10^{-2},10^{-1} \right\rbrace$.}
\label{fig:pullintimes}
\end{figure*}

In figure \ref{fig:pullintimes}a we compare the simulated pull-in times to the asymptotic prediction \eqref{eqn:bottlenecktime}, evaluated using the above values of $c_1$ and $c_2$. The asymptotic prediction provides an excellent approximation provided $Q \lesssim 10^{-2}$ and $\epsilon \lesssim 1$, with the numerics clearly following the predicted $\epsilon^{-1/2}$ scaling law. The accuracy of the asymptotics is remarkable: even though the result  \eqref{eqn:bottlenecktime} is based on our earlier assumption that $T_{\mathrm{PI}} \gg 1$, the computed times for $\epsilon \lesssim 1$ lie in the range $T_{\mathrm{PI}} \in (10^{-3},10^{-1})$. For values $Q \gtrsim 10^{-1}$, inertial effects are important when the beam reaches the pull-in displacement, and a bottleneck phase evidently does not occur (figure \ref{fig:pullintimes}a). Unfortunately, without an analytical solution of the dynamic beam equation \eqref{eqn:beam}, we are unable to predict a threshold value of $Q$ below which bottleneck behaviour occurs, since this requires knowledge of the solution before it reaches the fold shape. 

When we fix $\epsilon \lesssim 10^{-2} $, the pull-in time is a non-monotonic function of $Q$ (figure \ref{fig:pullintimes}a): pull-in occurs more quickly when $Q = 10^{-1}$ (green circles) compared to $Q = 10^{-2}$ (red diamonds), but is slower when $Q  = 1$ (magenta triangles). This feature is illustrated more clearly in figure \ref{fig:pullintimes}b, which shows a surface plot of the computed pull-in times as a function of $\epsilon$ and $Q$. In particular, when $\epsilon \lesssim 10^{-2}$ a minimum pull-in time is obtained when $Q \approx 0.04$. This minimum corresponds to a delicate balance between beam inertia and critical slowing down: inertia is large enough to prevent much slowing down in a bottleneck, but still small enough for the beam to be rapidly accelerated from its rest position. Very similar behaviour has been observed by \cite{gomez2018} for a parallel-plate capacitor (compare figure \ref{fig:pullintimes} here to figure $4$ in \cite{gomez2018}).

\begin{figure}
\centering
\includegraphics[width = \columnwidth]{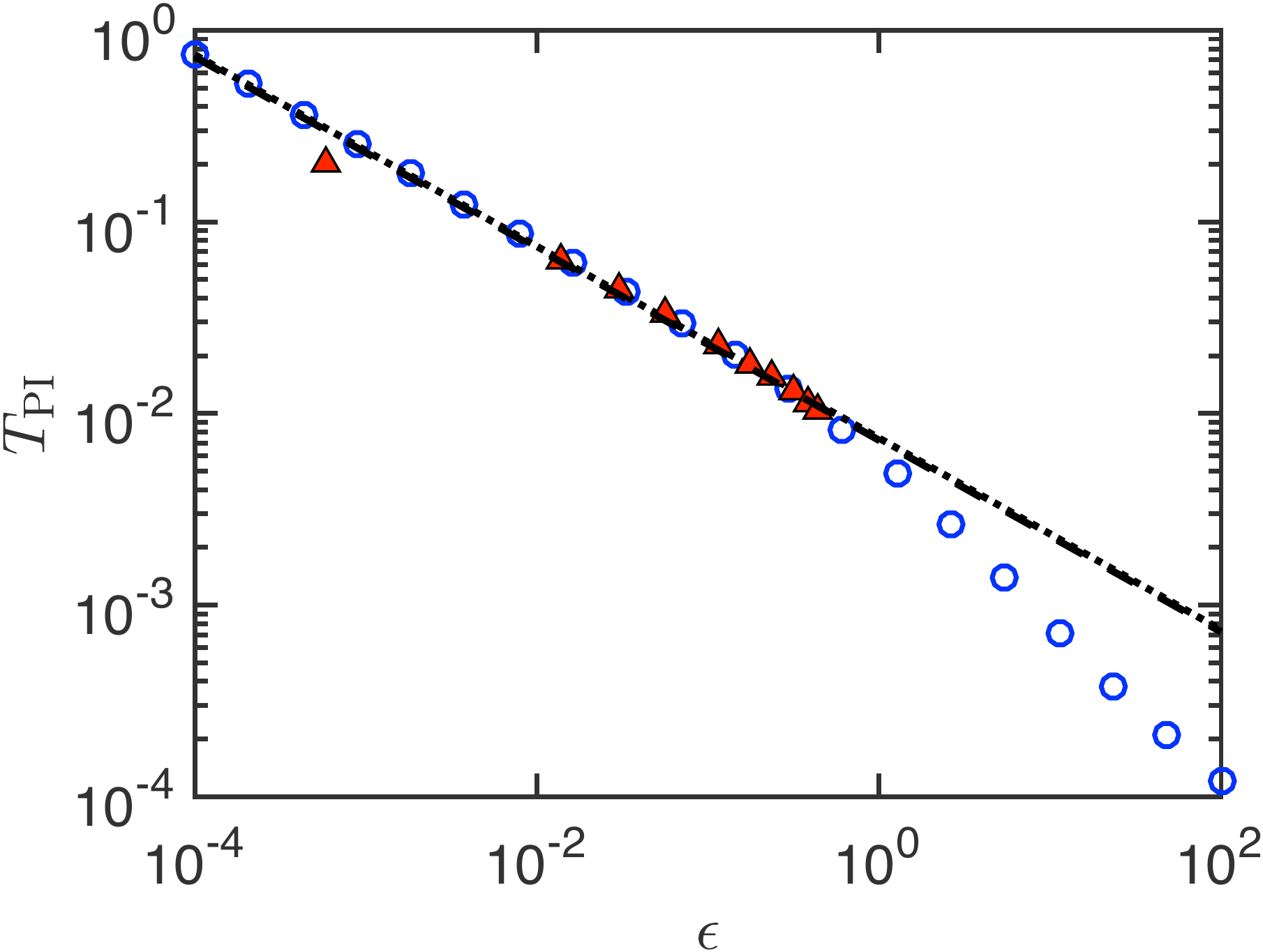} 
\caption{Pull-in times determined by previous simulations \cite{younis2003} using a reduced-order model (red triangles), and here using the method of lines with $N = 100$ grid points (blue circles) ($\cS \approx 0.06396$, $\tau_0 \approx -27.31$, $Q \approx 0.003846$). Also shown is the asymptotic prediction \eqref{eqn:bottlenecktime} from the bottleneck analysis (black dotted line), and, for later comparison, the prediction \eqref{eqn:SDOFpullintime} from the SDOF approximation (black dashed line).}
\label{fig:younisdata}
\end{figure}

To validate our numerics, we compare our results to numerical solutions reported by \cite{younis2003}, who solve the dynamic beam equation \eqref{eqn:beam} subject to \eqref{eqn:hookeslaw}--\eqref{eqn:ic} using a reduced-order model constructed by a Galerkin procedure (with the undamped eigenfunctions of the flat beam as basis functions). The parameter values in their study are  
$d_0 = 2.3~\mu\mathrm{m}$, $L = 610~\mu\mathrm{m}$, $h = 2.015~\mu\mathrm{m}$, $b = 40~\mu\mathrm{m}$, $E = 149~\mathrm{GPa}$, $\rho_s = 2.33~\mathrm{g}~\mathrm{cm}^{-3}$, $P_0/(bh) = -3.7~\mathrm{MPa}$, $V_{\mathrm{SPI}} = 8.76~\mathrm{V}$, and $\eta/\sqrt{\rho_s b h B/L^4} = Q^{-1} = 260$,
which correspond to
\beqn
[t] \approx 20.80~\mathrm{ms}, \quad \cS \approx 0.06396, \quad \tau_0 \approx -27.31.
\eeqn
We have extracted the pull-in times reported by \cite{younis2003}, as a function of the applied voltage $V$, using the WebPlotDigitizer (arohatgi.info/WebPlotDigitizer). We then use the reported pull-in voltage $V_{\mathrm{SPI}}$ to determine the corresponding values of $\epsilon= (V/V_{\mathrm{SPI}})^2 -1$, and non-dimensionalise the pull-in times using the overdamped timescale $[t]$. The results are in excellent agreement with our numerical simulations; see figure \ref{fig:younisdata}. (The discrepancy at the smallest value of $\epsilon$ is likely due to the error in extracting the point graphically using WebPlotDigitizer, or a possible rounding error in the reported pull-in voltage; either introduces a slight shift in the computed values of $\epsilon$, which is exaggerated for small values on log--log axes.) For the above parameter values we also compute
\beq 
\lamf \approx 38.0173 , \quad c_1 \approx 111.5, \quad c_2 \approx 1601. \label{eqn:younisvalues}
\eeq
The predicted pull-in time \eqref{eqn:bottlenecktime} is also plotted in figure \ref{fig:younisdata} (black dotted line) and fits well the numerical data without any adjustable parameters. 

\subsection{Comparison with other data}
We have shown that near the static pull-in transition, the dimensional pull-in time is 
\beq
t_{\mathrm{PI}} \sim \frac{L^4 \eta}{B}  \frac{\pi}{\sqrt{c_1 c_2 \epsilon}} \quad \mathrm{where} \quad \epsilon = \left(\frac{V}{V_{\mathrm{SPI}}}\right)^2-1. \label{eqn:pullintimedim}
\eeq
This result is valid for $0 < \epsilon \ll 1$ and $ Q \ll 1$. We note that the beam length $L$ and bending stiffness $B$ are quantities that are measurable in experiments. However, as discussed at the start of \S\ref{sec:formulation}, the damping coefficient $\eta$ is a lumped constant that parameterises the properties of the squeeze film, specifically during the bottleneck phase. We now show that this damping model, despite its simplicity, is able to approximate well experiments and numerical simulations of microbeams that incorporate compressible squeeze film damping.

\begin{table*}
\caption{Summary of data for the pull-in time of microbeams reported in the literature. \label{table:microbeamdata}}
\begin{indented} 
\item[]
\vspace{10pt}
    \begin{tabular}{m{0.3cm}m{1.3cm}m{0.4cm}m{0.4cm}m{0.4cm}m{0.4cm}m{0.4cm}m{0.4cm} m{0.7cm}m{0.6cm}m{2.3cm}m{1.1cm}m{1.6cm}m{0.9cm}}
       \hlineB{4}
  Ref. &  Data type & $d_0$ &  $L$ & $h$ & $b$ & $E$  & $V_{\mathrm{SPI}}$  & $\rho_s$ & $P_0/bh$ & Model &  Fitted $\eta$ & Estimated $\eta$ & Legend \\
   & &  $ (\mu\mathrm{m})$ &  $ (\mu\mathrm{m})$ & $ (\mu\mathrm{m})$ & $(\mu\mathrm{m})$ & $(\mathrm{GPa})$  & $(\mathrm{V})$  & $(\mathrm{g}\mathrm{cm}^{-3})$ & $(\mathrm{MPa})$ &  & $(\mathrm{Pa}~\mathrm{s})$ & $(\mathrm{Pa}~\mathrm{s})$ & \\ \hlineB{4}
           \cite{gupta1996} & Experiment & 2.07 & 610 &  2.12 & 40  & 164 & 8.76 & 2.2 & -3.5 & N/A & 0.802 &  1.10 &   \includegraphics[scale = 1]{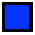} \\   
           \cite{gupta1996} & Experiment & 2.07 & 710 &  2.12 & 40 & 164 & 5.54 & 2.2 & -3.5  & N/A & 0.754 & 1.48 &  \includegraphics[scale = 1]{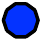} \\   
           \cite{hung1999} & Simulation & 2.3 & 610 & 2.2 & 40 & 149 & 8.76 & 2.33 & -3.7 & BE, CC, CSQFD, SBC & 0.613 & 0.908 &  \includegraphics[scale = 1]{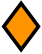} \\ 
            \cite{missoffe2008} & Simulation & 2.3 & 610 & 2.2 & 40 & 149 & 8.76 & 2.33 & -3.7 & BE, CC, CSQFD, $1.013~\mathrm{bar}^{\ddagger}$, SBC & 0.569 & 0.908 &  \includegraphics[scale = 1]{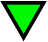} \\    
             \cite{missoffe2008} & Simulation & 2.3 & 610 & 2.2 & 40 & 149 & 8.76 & 2.33 & -3.7 & BE, CC, CSQFD, $0.1013~\mathrm{bar}^{\ddagger}$, SBC & 0.156 & 0.908 &  \includegraphics[scale = 1]{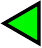} \\    
              \cite{missoffe2008} & Simulation & 2.3 & 610 & 2.2 & 70 & 149 & 8.76 & 2.33 & -3.7 & BE, CC, CSQFD, $0.1013~\mathrm{bar}^{\ddagger}$, SBC & 0.536 & 4.87 &  \includegraphics[scale = 1]{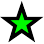} \\      
               \hlineB{4}
    \end{tabular}
    \vspace{5pt}
\item[] BE, dynamic beam equation; CC, clamped-clamped boundary conditions; CSQFD, compressible squeeze film damping (Reynolds equation); SBC, corrections due to slip boundary conditions (rarefaction effects); N/A, not applicable  
\item[] $^{\ddagger}$Ambient pressure   
\end{indented}
\end{table*}

\begin{figure}
\centering
\includegraphics[width = \columnwidth]{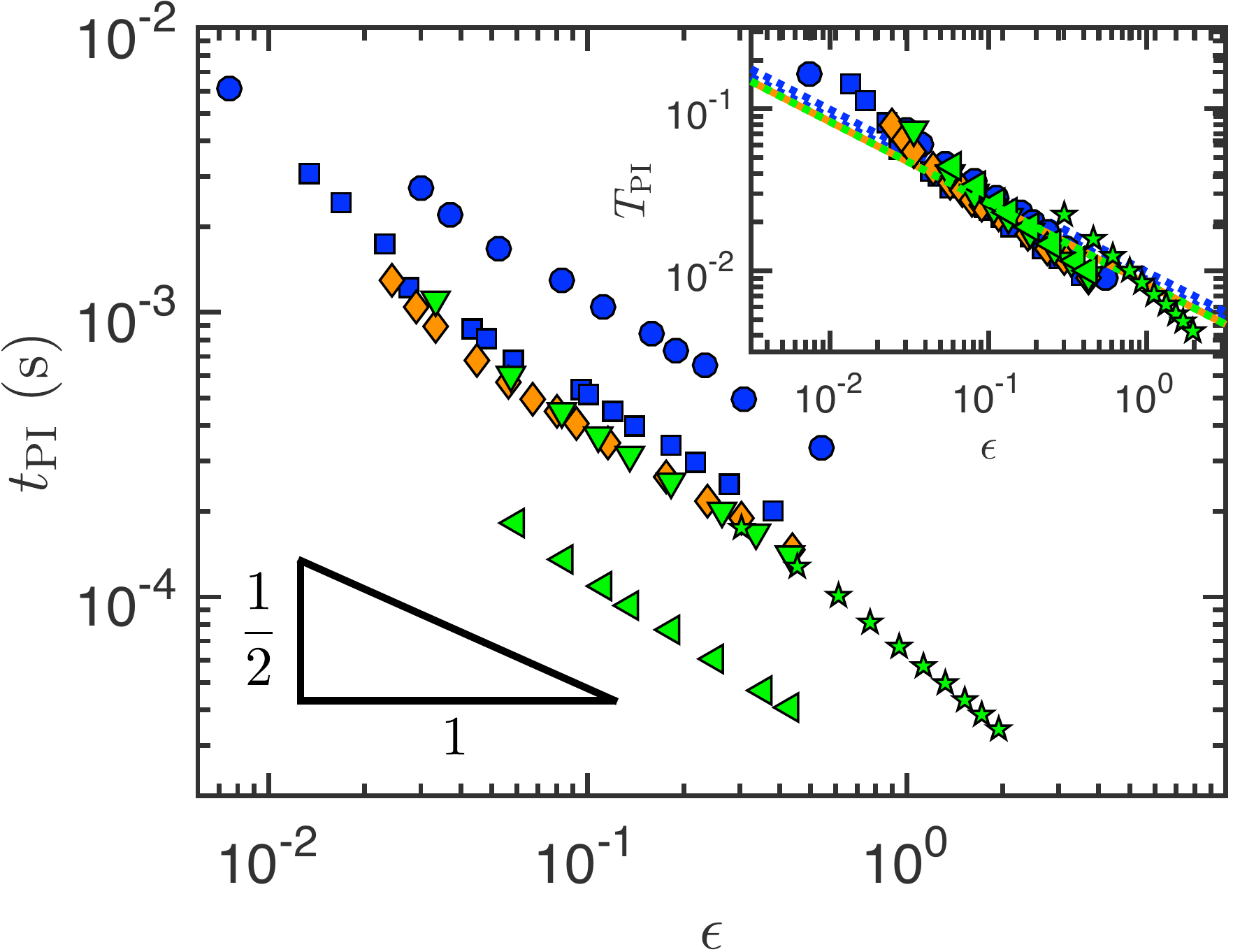} 
\caption{Main panel: Experimental and numerical pull-in times of overdamped microbeams reported in the literature, plotted as a function of the normalised voltage difference $\epsilon$. For a legend and the parameter values used  in each data set, see  table \ref{table:microbeamdata}. Inset: The same data made dimensionless using the overdamped timescale $[t] = L^4 \eta/B$; here $\eta$ is fitted to match the values with the asymptotic prediction \eqref{eqn:pullintimedim} (dotted lines).}
\label{fig:otherdata}
\end{figure}

We consider experiments performed by \cite{gupta1996} in air at atmospheric pressure. As in our model, the beams have clamped ends and are subject to step DC voltages. We also consider numerical simulations that model these experiments, which couple the dynamic beam equation to the compressible Reynolds equation in the squeeze film. The parameter values used in each study are summarised in table \ref{table:microbeamdata}. We have separated the data into rows so that within each data set only the actuation voltage changes: the properties of the beam and the  squeeze film do not vary. We also report any additional effects that are incorporated in the simulations. The dimensional pull-in times are plotted as a function of $\epsilon = (V/V_{\mathrm{SPI}})^2-1$ in figure \ref{fig:otherdata} (main panel). Here symbols are used to indicate different data sets, and colours are used to distinguish references (see the `Legend' column in table \ref{table:microbeamdata}). In all cases the slowing down near the pull-in transition approximately obeys an $\epsilon^{-1/2}$ scaling law. 

We make the pull-in times dimensionless using the following procedure. For each data set, we calculate the stretchability $\cS$ and residual tension $\tau_0$. By solving the corresponding neutral stability problem \eqref{eqn:neutralstability} at the saddle-node bifurcation, we numerically compute the dimensionless constants $c_1$ and $c_2$. We then compare the dimensional pull-in times with the prediction \eqref{eqn:pullintimedim}, and determine $\eta$ using a least-squares fit. The resulting pull-in times, made dimensionless using the overdamped timescale $[t] = L^4 \eta/B$, are shown in the inset of figure \ref{fig:otherdata}. In all cases we obtain good agreement with the dimensionless prediction \eqref{eqn:bottlenecktime} (dotted lines). Moreover, because the values of $\cS$ and $\tau_0$ are so similar between the data sets, the non-dimensionalisation also collapses the data well. 

We note that this fitting is very similar to that performed in \cite{gomez2018}, in which we collapsed a large range of data by fitting the overdamped timescale $[t]$ (see figure $6$ in \cite{gomez2018}). However, the approach here has the advantage that the parameters of the beam are explicitly accounted for in the timescale $[t]$ and constants $c_1$ and $c_2$. Hence, once the damping coefficient has been fitted for one data set, it is possible  to use equation \eqref{eqn:pullintimedim} to make further predictions if the parameters of the beam then change. For example, if the residual tension is varied, it is only necessary to compute the updated values of $c_1$ and $c_2$. 

We also check that the fitted values of $\eta$ are realistic by comparing to an approximate analytical solution. Because the beam is shallow, when viewed on the length scale of the squeeze film, it approximately acts as an infinitely long and flat rectangular plate that moves in the perpendicular direction only. The incompressible Reynolds equation may be solved approximately in this geometry  \citep{krylov2004} to give the damping coefficient $\eta \approx \mu b^3/d^3$, where $\mu$ is the air viscosity and $d$ is the local gap width. For the microbeam considered here, the dimensionless displacement during the bottleneck phase is $W \approx \Wf$. In dimensional terms, the gap width is therefore $d_0[1- \Wf(X)]$, giving the damping coefficient 
\beqn
\eta(X) \approx \frac{\mu b^3}{d_0^3[1-\Wf(X)]^3}.
\eeqn
This varies along the length of the beam so we cannot compare it directly to our fitted values. As the minimum gap width is attained at $X = 1/2$ (figure \ref{fig:shapes}), an upper bound on the damping coefficient is 
\beq
\eta \lesssim \frac{\mu b^3}{d_0^3[1-\Wf(1/2)]^3}. \label{eqn:predictedb0}
\eeq
(Averaging $\eta(X)$ over the beam length instead does not give a useful estimate.) In table \ref{table:microbeamdata} we compare this prediction to the values obtained by fitting the pull-in times (for all data sets the air viscosity $\mu = 18.2~\mu\mathrm{Pa}~\mathrm{s}$). The values are of comparable size for all data sets, with \eqref{eqn:predictedb0} indeed providing an upper bound. The fitting here is therefore consistent with the damping being dominated by viscous dissipation rather than air compressibility. The discrepancy between the values may also be due to additional effects present in the experiments and numerical simulations, which are not captured by the expression \eqref{eqn:predictedb0}. These include finite-length effects (i.e.~venting conditions at the clamped boundaries) and rarefaction effects. (This explains why the discrepancy is largest for the data of \cite{missoffe2008} at reduced ambient pressure $0.1013~\mathrm{bar}$, i.e.~the final two rows in table \ref{table:microbeamdata}; here we expect rarefaction effects to be more significant.) In experiments, material damping may also be present. Finally, we also note that with the fitted values of $\eta$, the corresponding quality factors $Q$ are all small compared to unity, consistent with our assumption that the microbeams are overdamped.

\section{Single-mode approximation}
\label{sec:singlemode}
In this final section, we consider the error in the pull-in time calculated using a standard SDOF approximation. We assume \emph{a priori} that the displacement can be written in the separable form
\beq
W(X,T) = U(T)\Phi(X), \label{eqn:singlemode}
\eeq
where $U(T)$ is an unknown amplitude and $\Phi(X)$ is a known spatial function. We focus on the commonly used choice of $\Phi(X)$ as the first eigenfunction of the flat beam, i.e.~the fundamental vibrational mode when the applied voltage is zero; this is computed in Appendix C. In this way, equation \eqref{eqn:singlemode} may be interpreted as keeping only the first term in a standard Galerkin expansion that uses these eigenfunctions as basis functions  \citep{younis2003,krylov2007,joglekar2011}. 
We insert the separated ansatz \eqref{eqn:singlemode} into the dynamic beam equation \eqref{eqn:beam} to obtain 
\beq
\Phi \left(Q^2 \dd{U}{T}+\d{U}{T}\right)+\left(\df{\Phi}{X}-\tau\dd{\Phi}{X}\right)U = \frac{\lambda}{\left(1-\Phi U\right)^2}. \label{eqn:beamsinglemode}
\eeq
The Hooke's law constraint \eqref{eqn:hookeslaw} becomes
\beqn
\tau = \tau_0 + \frac{U^2}{2\cS}\int_0^1 \left(\d{\Phi}{X}\right)^2~\id X.
\eeqn
Combining this with the ODE satisfied by $\Phi(X)$ (equation \eqref{eqn:beamstabilitysmallvoltage} in Appendix C), the spatial derivatives appearing in the beam equation \eqref{eqn:beamsinglemode} can be written as
\beqn
\df{\Phi}{X}-\tau\dd{\Phi}{X} = \Omega^2 \Phi - \frac{U^2}{2\cS}\dd{\Phi}{X}\int_0^1 \left(\d{\Phi}{X}\right)^2~\id X,
\eeqn
where $\Omega$ is related to the natural frequency of the beam. Multiplying \eqref{eqn:beamsinglemode} by $\Phi$ and integrating from $X = 0$ to $X = 1$ (simplifying using integration by parts and the boundary conditions/normalisation satisfied by $\Phi$; see Appendix C), we obtain an ODE for $U$:
\begin{eqnarray}
Q^2 \dd{U}{T}+\d{U}{T} + \Omega^2 U & + & \frac{1}{2\cS}\left[\int_0^1 \left(\d{\Phi}{X}\right)^2~\id X\right]^2 U^3 \nonumber \\ 
& = & \lambda \int_0^1 \frac{\Phi}{\left(1-\Phi U \right)^2}~\id X. \label{eqn:beamsinglemodeODE}
\end{eqnarray}
It is not clear how to write the integral on the right-hand side as a simple function of $U$. While this could be avoided by multiplying equation \eqref{eqn:beamsinglemode} by $(1-\Phi U)^2$ before integrating, the form here is more convenient and makes the physical nature of each term apparent. In particular, the linear term on the left-hand side represents the effective spring force due to the bending stiffness of the beam, while the cubic correction represents additional stiffening due to stretching (strain-stiffening). 

\subsection{Steady solutions}
Steady solutions of \eqref{eqn:beamsinglemodeODE} satisfy
\beq 
\lambda = \frac{\Omega^2 U + \left[\int_0^1 \left(\id \Phi/\id X\right)^2~\id X\right]^2 U^3/(2\cS)}{\int_0^1\Phi \left(1-\Phi U \right)^{-2}~\id X}. \label{SDOFsteadysoln}
\eeq
For given $\cS$ and $\tau_0$, the above relation allows us to compute the corresponding values of $\lambda$ as $U$ varies (using quadrature to evaluate the integrals). The midpoint displacement is given in terms of $U$ by
\beqn
W(1/2) = \Phi(1/2)U.
\eeqn
Response diagrams of $W(1/2)$ as a function of $\lambda$ obtained in this way are superimposed (as black dashed curves) on figures \ref{fig:response}a--b. These show that the SDOF method provides a remarkably good approximation of the numerically computed bifurcation diagrams. The disagreement is largest in the neighbourhood of the fold point, where the solution becomes highly sensitive to changes in $\lambda$; similar behaviour has been reported by \cite{younis2003} and \cite{krylov2008}. 

We write $\lamf^{\mathrm{SDOF}}$ for the value of $\lambda$ at the fold, which corresponds to $U = \Uf$ in this approximation; the $\mathrm{SDOF}$ superscript on $\lambda$ is to distinguish its value from that obtained in \S\ref{sec:equilibria}, when we solved the full beam equation using \texttt{bvp4c}. We now obtain two identities that will be useful in the dynamic analysis. Because the fold point is a steady solution, we have from \eqref{SDOFsteadysoln}
\beq
\lamf^{\mathrm{SDOF}} = \frac{\Omega^2 \Uf + \left[\int_0^1 \left(\id \Phi/\id X\right)^2~\id X\right]^2 \Uf^3/(2\cS)}{\int_0^1\Phi \left(1-\Phi \Uf \right)^{-2}~\id X}. \label{eqn:singlemodefoldeqn1}
\eeq
In addition, the fact that this is a fold gives that $(\partial \lambda/\partial U)|_{U = \Uf,~\lambda = \lamf^{\mathrm{SDOF}}} = 0$ which, using \eqref{eqn:singlemodefoldeqn1}, can be simplified to
\begin{eqnarray}
\Omega^2 + \frac{3\Uf^2}{2\cS}\left[\int_0^1 \left(\d{\Phi}{X}\right)^2~\id X\right]^2 \nonumber \\ 
 \qquad \qquad \quad = 2\lamf^{\mathrm{SDOF}} \int_0^1 \frac{\Phi^2}{\left(1-\Phi \Uf \right)^3}~\id X.   \label{eqn:singlemodefoldeqn2}
\end{eqnarray}

\subsection{Pull-in dynamics}
\label{sec:SDOFdynamics}
Using the SDOF approximation, we would like to calculate the pull-in time when
\beqn
\lambda = \lamf^{\mathrm{SDOF}} (1+\epsilon),
\eeqn 
for $0 < \epsilon \ll 1$. From \S\ref{sec:dynamics}, we know that the dynamics are highly sensitive in this regime, with a small change in $\epsilon$ producing a large change in pull-in time. Because of the error between $\lamf^{\mathrm{SDOF}}$ and the `true' bifurcation value $\lamf$ (i.e.~from solving the full beam model without making a SDOF approximation), replacing $\lamf^{\mathrm{SDOF}}$ by $\lamf$ above will lead to large errors in the pull-in time: any difference in estimates of $\lamf$ changes the effective value of $\epsilon$. (We also discuss this sensitivity in Appendix A in the context of solving the PDE numerically.)  A similar issue has been described by \cite{joglekar2011} in their analysis of an underdamped microbeam, who found that the error in the SDOF approximation is very large at voltages near the dynamic pull-in voltage. We now show that it is possible to obtain excellent agreement when using a SDOF approach, provided one uses the bifurcation value $\lamf^{\mathrm{SDOF}}$, i.e.~the value consistent with the SDOF equations. 

When the solution is close to the pull-in displacement we have
\beqn
U(T) = \Uf + \Ut(T),
\eeqn
where $|\Ut| \ll 1$. We expand the electrostatic force in \eqref{eqn:beamsinglemodeODE} as
\begin{eqnarray*}
\lambda \int_0^1 \frac{\Phi}{\left(1-\Phi U \right)^2}~\id X & = &  \lamf^{\mathrm{SDOF}}(1+\epsilon)I_1 + 2\lamf^{\mathrm{SDOF}}I_2 \Ut \\
&& \: + 3\lamf^{\mathrm{SDOF}} I_3 \Ut^2 +  O(\epsilon\Ut,\Ut^3),
\end{eqnarray*}
where we define
\beqn
I_m = I_m(\Phi,\Uf) =  \int_0^1 \frac{\Phi^m}{(1-\Phi\Uf)^{m+1}}\:\id X.
\eeqn
We substitute into \eqref{eqn:beamsinglemodeODE} and simplify using the identities \eqref{eqn:singlemodefoldeqn1}--\eqref{eqn:singlemodefoldeqn2}. Neglecting the inertia term and terms of $O(\epsilon\Ut,\Ut^3)$, we obtain at leading order
\beqn
\d{\Ut}{T} = d_1 \epsilon + d_2 \Ut^2, 
\eeqn
where 
\begin{eqnarray*}
d_1 & = & \lamf^{\mathrm{SDOF}} I_1, \\
d_2 & = & 3\lamf^{\mathrm{SDOF}} I_3 - \frac{3\Uf}{2\cS}\left[\int_0^1 \left(\d{\Phi}{X}\right)^2~\id X\right]^2.
\end{eqnarray*}
These equations are precisely equations \eqref{eqn:bottleneckode}--\eqref{eqn:defnc1c2}, derived in the bottleneck analysis of the full PDE in \S\ref{sec:dynamics}, provided that we identify
\begin{eqnarray}
A \to \Ut, \quad (c_1,c_2) \to (d_1,d_2),  \quad \lamf \to \lamf^{\mathrm{SDOF}}, \nonumber \\
W_p \to \Phi, \quad  \Wf \to \Uf \Phi. \label{eqn:SDOFanalogy}
\end{eqnarray}
The pull-in time to leading order is then similarly evaluated as
\beq
T_{\mathrm{PI}} \sim \frac{\pi}{\sqrt{d_1 d_2 \epsilon}}, \label{eqn:SDOFpullintime}
\eeq
and the displacement in the bottleneck is
\beq
W(X,T) = \Uf\Phi(X) + \sqrt{\frac{d_1 \epsilon}{d_2}} \Phi(X) \tan\Big[\sqrt{d_1 d_2 \epsilon}~T-\frac{\pi}{2}\Big].
\label{eqn:SDOFdisplacement}
\eeq

This analogy with our analysis of the PDE model may not be so unexpected. In \S\ref{sec:dynamics} we first expanded the solution about the equilibrium shape at the fold, before performing a SDOF-like approximation (using the neutrally-stable eigenfunction $W_p$   as a basis function). In this section we essentially performed these steps in the reverse order: we first used a SDOF approximation to reduce the beam equation to an ODE, and then expanded about the fold solution. The analogy in \eqref{eqn:SDOFanalogy} also allows us to deduce that the error in the second approach is governed by three quantities. These are (i) the error between $\lamf$ and $\lamf^{\mathrm{SDOF}}$; (ii) the error between $W_p$ and the basis function $\Phi$, here chosen as the eigenfunction of the flat beam; and (iii) the error between the fold shape $\Wf$ and the approximation $\Uf \Phi$. Together, these errors govern the difference between the constants $(c_1,c_2)$ and $(d_1,d_2)$ and hence the discrepancy in the predicted pull-in time and bottleneck displacement. 


To quantify these errors, we again consider the case $\cS = 10^{-2}$ and $\tau_0 = 0$. For the SDOF system we calculate 
\begin{eqnarray*}
\lamf^{\mathrm{SDOF}} \approx 179.9184, \quad \Uf\Phi(1/2) \approx 0.6123, \\
\Phi(1/2) \approx 1.588, \quad d_1 \approx 626.6, \quad d_2 \approx 11135. 
\end{eqnarray*}
These values agree well with the corresponding quantities obtained in \S\ref{sec:dynamics}; compare to equation \eqref{eqn:S1e-2tau0values}. In figures \ref{fig:trajectories}a--b, we have superimposed the midpoint displacement predicted by equation \eqref{eqn:SDOFdisplacement} (as black dashed curves), which consistently under-predicts the duration of the bottleneck phase compared to the bottleneck analysis of the PDE model. Nevertheless, the SDOF approximation closely captures the dependence of the pull-in time on $\epsilon$ --- the relative error in the pre-factor of $\epsilon^{-1/2}$ between the two approaches is around $7\%$.  This is evident in figure \ref{fig:pullintimes}a, where the predictions of the bottleneck analysis and SDOF approximation are almost indistinguishable.

A similar picture is seen for the parameter values used by \cite{younis2003}. Here we calculate for the SDOF system
\begin{eqnarray*}
\lamf^{\mathrm{SDOF}} \approx 38.3153, \quad \Uf\Phi(1/2) \approx 0.5720, \\
\Phi(1/2) \approx 1.6165, \quad d_1 \approx 112.9, \quad d_2 \approx 1664,
\end{eqnarray*}
which closely match the quantities \eqref{eqn:younisvalues} obtained in \S\ref{sec:dynamics}. Again, we find that the pre-factors are in very good agreement between the two approaches, with a relative error of around $3\%$ in this case (figure \ref{fig:younisdata}). 

\section{Discussion and conclusions}
\label{sec:conclusions}
In this paper, we have analysed the pull-in dynamics of overdamped microbeams. Rather than using a one-dimensional mass-spring model, we explicitly considered the beam geometry, modelled using the dynamic beam equation. Using direct numerical solutions, we demonstrated that at voltages just beyond the pull-in voltage, the dynamics slow down considerably in a bottleneck phase. This phase is similar to the `metastable' interval first described by \cite{rocha2004a} for a parallel-plate actuator and analysed in detail \citep{gomez2018}: the bottleneck depends sensitively on the voltage, it dominates the time taken to pull-in, and occurs when the solution passes the static pull-in displacement.

We analysed the bottleneck dynamics using two approaches. In the first approach we worked with the dynamic beam equation directly. Because a linear stability analysis is not applicable (there is no unstable base state from which the system evolves), we used an asymptotic method based on two quantities: the proximity of the solution to the pull-in displacement, and the slow bottleneck timescale. This allowed us to systematically reduce the leading-order dynamics to a simple amplitude equation --- the normal form for a saddle-node bifurcation. As a result, the microbeam dynamics inherit the critical slowing down due to the `ghost' of the saddle-node bifurcation \citep{strogatz}. We obtained a simple approximation to the pull-in time: 
\beqn
t_{\mathrm{PI}} \sim \frac{L^4 \eta}{B}  \frac{\pi}{\sqrt{c_1 c_2 \epsilon}},
\eeqn
where $L$ is the beam length, $\eta$ is the effective damping coefficient during the bottleneck phase, $B$ is the bending stiffness, $c_1$ and $c_2$ are dimensionless constants, and $\epsilon$ is the normalised difference between the applied voltage and the pull-in voltage. To compute $c_1$ and $c_2$ requires some effort: it is necessary to solve for the equilibrium shape at the fold, $\Wf(X)$ (e.g.~using a continuation algorithm), as well as the neutrally-stable eigenfunction of the fold shape, $W_p(X)$. These problems depend on the beam stretchability, residual stress and the boundary conditions applied to the beam.


In the second approach, we first applied a SDOF approximation, assuming that the solution can be written in the separable form $W(X,T)=U(T)\Phi(X)$. By reducing the dynamic beam equation to an ODE, we were able to analyse the behaviour near the pull-in transition in a similar way to \cite{gomez2018}. Comparing this approach to the bottleneck analysis of the PDE model revealed that three factors control the error of the SDOF approximation: 
\begin{enumerate}[label=(\roman*)]
\item{The error in the computed pull-in voltage.}
\item{The error between the basis function $\Phi(X)$ and the neutrally stable eigenfunction $W_p(X)$.}
\item{The error in the computed pull-in displacement.}
\end{enumerate}
We found that choosing $\Phi(X)$ to be the  fundamental vibrational mode of the undeformed beam closely matches $W_p(X)$ (the same eigenfunction when evaluated at the pull-in voltage), so that the error (ii) is small. Moreover, it may be verified that $\Phi(X)$ is left-right symmetric about the beam midpoint $X = 1/2$. Because the pull-in displacement is also left-right symmetric, it turns out that the errors (i) and (iii) above are also small. As a consequence, we could obtain accurate predictions for the pull-in time with much less effort. This result is in direct contrast to previous studies, which conclude that the error in the SDOF approximation grows unacceptably large near pull-in \citep{joglekar2011} -- the apparent discrepancy is because our approach accounts for the shift in the pull-in voltage when using the SDOF system, and so is consistent with its bifurcation behaviour (recall the discussion at the start of \S\ref{sec:SDOFdynamics}). However, in other scenarios (e.g.~different boundary conditions) it is possible that the errors (i)--(iii) could be large, meaning the SDOF approximation is no longer valid. In such cases $W_p$ should instead be used as the basis function, provided that the error in the pull-in displacement is verified to be small.

Finally, we discuss the various assumptions we have made in our analysis. We assumed that the quality factor $Q$ is small, so that inertial effects can be neglected. This is necessary to obtain bottleneck behaviour near the static pull-in transition when the voltage is stepped from zero (for large $Q$ the beam is not slowed in a bottleneck due to inertia). We focussed on the case of a clamped-clamped beam under step DC loads, though our analysis may be adapted to other boundary conditions and loading types. However, we note that in the case of a voltage sweep, dynamic effects may cause a delayed bifurcation and modify the $\epsilon^{-1/2}$ scaling law governing the bottleneck duration, similar to what is observed in other physical systems \citep{tredicce2004,majumdar2013}. In addition, we neglected spatial variations in the damping coefficient, and used a lumped constant in our study. Because the bottleneck dominates the transient dynamics, and the beam geometry is roughly constant in the bottleneck, we found that this is sufficient to accurately predict the pull-in time --- we were able to collapse data from experiments and simulations that incorporate compressible squeeze film damping (figure \ref{fig:otherdata}). Nevertheless, the framework we have presented here shows that it is the underlying bifurcation structure that governs the bottleneck dynamics, so that more realistic damping models could also be incorporated. 

\paragraph{Acknowledgments}

The research leading to these results has received funding from the European Research Council under the European Union's Horizon 2020 Programme/ERC Grant No. 637334 (D.V.) and the EPSRC Grant No. EP/ M50659X/1 (M.G.). The data that supports the plots within this paper and other findings of this study are available from https://doi.org/10.5287/bodleian:QmowO9Q06.

\appendix
\setcounter{section}{0}

\section{Details of the numerical scheme}
To solve the dynamic beam equation, we introduce a uniform mesh on the interval $[0,1]$ with spacing $\Delta X = 1/N$, where $N\geq 2$ is an integer. We label the grid points as $X_i = i\Delta X$ ($i = 0,1,2,\ldots,N$) and write $W_i$ for the numerical approximation of $W$ at the grid point $X_i$. We approximate the spatial derivatives appearing in the beam equation \eqref{eqn:beam} and boundary conditions  \eqref{eqn:bc} using centered differences with second-order accuracy. To do this at all interior points in the mesh without losing accuracy, we introduce the ghost points $X_{-1}$ and $X_{N+1}$ (with associated displacement $W_{-1}$ and $W_{N+1}$) outside of the interval $[0,1]$. With this scheme,  \eqref{eqn:beam} becomes, for $i = 1,2,\ldots,N-1$,
\begin{eqnarray}
 &&Q^2\dd{W_i}{T}+\d{W_i}{T} - \tau \frac{W_{i+1}-2W_{i}+W_{i-1}}{\Delta X^2} \nonumber \\
&&+ \frac{W_{i+2}-4 W_{i+1}+6 W_{i}-4 W_{i-1}+W_{i-2}}{\Delta X^4} = \frac{\lambda}{(1-W_i)^2}. \nonumber \\ \label{eqn:discretisedbeam}
\end{eqnarray}
The clamped boundary conditions \eqref{eqn:bc} are approximated by
\beq
W_0 = \frac{W_1-W_{-1}}{2\Delta X} = W_N = \frac{W_{N+1}-W_{N-1}}{2\Delta X} = 0.
\label{eqn:discretisedclamped}
\eeq
To approximate the integral appearing in the Hooke's law constraint  \eqref{eqn:hookeslaw}, we use a centered difference to discretise the derivative and apply the trapezium rule for the quadrature. After simplifying using the clamped conditions, we obtain
\beq
\mathcal{S}(\tau-\tau_0) = \frac{1}{8\Delta X}\sum_{k=1}^{N-1}\left(W_{k+1}-W_{k-1}\right)^2. \label{eqn:discretisedHooke}
\eeq
Finally, we have the initial data $W_i(0) = \dot{W}_i(0) = 0$ ($i = 1,2,\ldots,N-1$).


These equations can readily be written in matrix form and integrated using the \textsc{matlab} ODE solvers. 
We use the routine \texttt{ode23t}, which employs a stiff solver to efficiently integrate the system when $Q \ll 1$; here the equations are stiff due to transients around $T = 0$ and immediately before contact in which inertia cannot be neglected. We have verified that (i) the equilibrium solutions of the discretised system converge to the solution of the steady beam equation (obtained using  \texttt{bvp4c}); and (ii) the solutions of the unsteady discretised equations, integrated up to a fixed time, converge as $N$ increases. In both cases, we observe second-order accuracy in the convergence; for further details and convergence plots see \cite{gomezthesis}.


When we set $\lambda = \lamf(1+\epsilon)$, we anticipate that the dynamics depend sensitively on the value of $\epsilon$ as $\epsilon \to 0$. An important point is that due to the discretisation error in our numerical scheme, there will also be an error in the bifurcation value $\lamf$. If we use the value of $\lamf$ predicted by the solution of the `continuous' problem (i.e.~from solving the beam equation using \texttt{bvp4c}) in our simulations, we therefore need to ensure that the relative error in $\lamf$ is much smaller than $\epsilon$: this error acts as an `extra' perturbation that shortens the pull-in time. For example, we find that taking $N = 600$ ensures a relative error that is typically $O(10^{-5})$, which is sufficient provided we restrict to $\epsilon \gtrsim 10^{-3}$.

An alternative approach is to use the value of $\lamf$ predicted from the discretised system, which is consistent with its bifurcation behaviour and so eliminates this sensitivity. This allows us to use fewer grid points, e.g.~$N = 100$, to obtain quantitatively similar results with much  less computing time. We use this latter approach for all simulations reported in this paper. To determine the value of $\lamf$ for the discretised system, we solve the steady version of equations \eqref{eqn:discretisedbeam}--\eqref{eqn:discretisedHooke} in \textsc{matlab} using the \texttt{fsolve} routine (error tolerances $10^{-10}$), using a simple continuation algorithm to trace the bifurcation diagram. Similar to the way we solved the steady beam equation in \S\ref{sec:equilibria}, we control the tension $\tau$ and determine $\lambda$ as part of the solution.

\section{Bottleneck analysis when $A = O(\epsilon)$}
In this appendix we show that the solution  \eqref{eqn:bottlenecksolnA}  for the amplitude variable $A(T)$ also holds when $A = O(\epsilon)$; using the expansion $\tan x \sim x$ for $|x| \ll 1$, this corresponds to times $|T - T_0| = O(1)$. Because the leading-order solution in the bottleneck is $O(\epsilon)$, our original assumption (ii) (made at the start of \S\ref{sec:bottleneckanalysis}) is no longer valid. Returning to the beam equation \eqref{eqn:bottleneckbeam}, we see that the left-hand side no longer dominates when $(\Wt_0,\taut_0) = O(\epsilon)$ and $|T - T_0| = O(1)$. We must now keep the time derivative and the $O(\epsilon)$ term on the right-hand side to obtain
\beq
L(\Wt_0,\taut_0) = -\pd{\Wt_0}{T} + \frac{\lamf}{(1-\Wf)^2}\epsilon. \label{eqn:bottleneckbeamsmallA}
\eeq
Because $|\Wt_0| \ll 1$, the Hooke's law constraint \eqref{eqn:bottleneckhookeslaw} remains unchanged at leading order:
\beqn
\cS \taut_0 = \int_0^1 \d{\Wf}{X}\pd{\Wt_0}{X}~\id X.
\eeqn
Again, we have an inhomogeneous problem and so the Fredholm Alternative Theorem applies. Integrating by parts (making use of the clamped boundary conditions and the Hooke's law constraints satisfied by $W_p$ and $\Wt_0$) shows that
\beqn
\int_0^1 W_p L(\Wt_0,\taut_0)~\id X = 0.
\eeqn
(This follows more generally from the fact that the operator $L(\cdot,\cdot)$, defined in  \eqref{eqn:bottleneckbeam}, is self-adjoint and the boundary conditions/constraints satisfied by $(W_p,\tau_p)$ and $(\Wt_0,\taut_0)$ are all homogeneous.) Multiplying \eqref{eqn:bottleneckbeamsmallA} by $W_p$ and  integrating over $(0,1)$ then gives
\beqn
0 = -\int_0^1 W_p \pd{\Wt_0}{T}~\id X + \epsilon \lamf \int_0^1 \frac{W_p}{(1-\Wf)^2}~\id X.
\eeqn
From the normalisation $\int_0^1 W_p^2\:\id X = 1$ (recall equation \eqref{eqn:neutralstability}), it follows that $\Wt_0 = A(T) W_p$ with 
\beqn
\d{A}{T} = \epsilon \lamf \int_0^1 \frac{W_p}{(1-\Wf)^2}~\id X = c_1 \epsilon.
\eeqn
The solution is
\beqn
A(T) = c_1\epsilon (T - T_0),
\eeqn
where the constant of integration is chosen to match into the solution \eqref{eqn:bottlenecksolnA} when $A \gg \epsilon$. This solution is precisely \eqref{eqn:bottlenecksolnA} when we expand the $\mathrm{tan}$ function for small arguments. We deduce that  \eqref{eqn:bottlenecksolnA} is asymptotically valid for all $|A| \ll 1$. 

\section{Linear stability at zero voltage}
In this appendix we determine the small-amplitude (flexural) vibrational modes of the flat beam when $\lambda = 0$, and the tension is close to the residual value ($\tau \approx \tau_0$). We set $W = \delta \Phi(X)e^{i\omega T}$ where $\delta \ll 1$ is a fixed quantity and $\omega$ is the (unknown) natural frequency. Inserting into the dynamic beam equation \eqref{eqn:beam} and considering terms of $O(\delta)$, we obtain \citep{younis2003} (assuming the real part of complex quantities) 
\beq
\df{\Phi}{X} - \tau_0 \dd{\Phi}{X} - \Omega^2 \Phi = 0, \label{eqn:beamstabilitysmallvoltage}
\eeq
where $\Omega^2 = Q^2\omega^2-i\omega$. In the absence of any damping, i.e.~as $Q \to \infty$, we see that $\Omega$ is simply proportional to the natural frequency of the beam. The clamped boundary conditions \eqref{eqn:bc} imply that $\Phi(0)=\Phi'(0)=\Phi(1)=\Phi'(1)=0$. The solution to \eqref{eqn:beamstabilitysmallvoltage} satisfying the boundary conditions at $X = 0$ is
\begin{eqnarray}
\Phi & = &  A_1\Big(\cosh\alpha_{+}X-\cos\alpha_{-}X\Big) \nonumber \\
&& \: + A_2\Big(\alpha_{-}\sinh\alpha_{+}X-\alpha_{+}\sin\alpha_{-}X\Big), \label{eqn:smalllambdaeigenmode}
\end{eqnarray}
where $A_1$ and $A_2$ are constants and we have introduced
\beqn
\alpha_{\pm} = \sqrt{\sqrt{\left( \frac{\tau_0}{2}\right)^2+\Omega^2}\pm \frac{\tau_0}{2}}.
\eeqn
The remaining boundary conditions, at $X = 1$, then yield
a second-order, homogeneous linear system in the two unknowns $A_1$ and $A_2$. To determine eigenfunctions, we are only interested in non-trivial solutions. These exist if and only if the corresponding determinant vanishes, which can be re-arranged to \cite{neukirch2012}
\beqn
\frac{\tau_0}{2 \Omega} = \frac{\cosh\alpha_{+}\cos\alpha_{-}-1}{\sinh\alpha_{+}\sin\alpha_{-}}.
\eeqn
For each value of $\tau_0$, the roots of this transcendental equation give the eigenvalues $\Omega$. The smallest positive root then corresponds to the fundamental eigenfunction $\Phi(X)$ used in \S\ref{sec:singlemode}; we specify the normalisation condition $\int_0^1\Phi^2\:\id X = 1$.

\end{document}